\documentclass[12pt]{article}
\usepackage[utf8]{inputenc}

\usepackage{courier}
\usepackage[margin=1in]{geometry}
\usepackage{graphicx}
\usepackage{natbib}
\usepackage{apalike}
\usepackage[hyphens]{url} %
\usepackage{amsmath}
\usepackage{amsfonts}
\usepackage{amssymb}
\usepackage{algpseudocode}
\usepackage{algorithm}
\usepackage{url}
\usepackage{subcaption}
\usepackage{xcolor}
\usepackage{multirow}
\usepackage{dsfont}
\usepackage{bm}
\usepackage{setspace}
\usepackage[title]{appendix}
\usepackage{authblk}
\usepackage{booktabs}

\providecommand{\keywords}[1]
{
  \small	
  \textbf{\textit{Keywords---}} #1
}

\providecommand{\botrule}
{
\bottomrule
}

\title{Detecting interpolation errors in infant mortality counts in 20th Century England and Wales}

\author[1]{Tessa Wilkie}
\author[2]{Idris Eckley}
\author[2,$\ast$]{Paul Fearnhead}
\author[3]{Ian Gregory}

\affil[1]{School of Computing and Communications, Lancaster University}
\affil[2]{School of Mathematical Sciences, Lancaster University}
\affil[3]{School of Global Affairs, Lancaster University}

\begin{document}

\newif\ifOUP
\OUPfalse

\newcommand{\MyAbstract}{Understanding historical datasets, such as the England and Wales infant mortality data for local government districts, can provide valuable insights into our changing society. Such analyses can prove challenging in practice, due to frequent changes in the boundaries of local government districts for which records are collected. One solution adopted in the literature to overcome such practical challenges is to pre-process data using areal interpolation to render the units consistent over the time period of focus. However, such methods are prone to errors. In this paper we introduce a novel changepoint method to detect instances where interpolation performs poorly. We demonstrate the utility of our method on original data, and also demonstrate how correcting interpolation errors can affect the clustering of the infant mortality curves. }

\newcommand{\MyKeywords}{changepoint, clustering, functional data, infant mortality, social history}

\newcommand{\AbstractWithKeywords}{
  \begin{abstract}
    \MyAbstract
  \end{abstract}
  \keywords{\MyKeywords}
}



\ifOUP
  \AbstractWithKeywords
\fi

\maketitle

\ifOUP\else
  \AbstractWithKeywords
\fi

\section{Introduction} \label{sec: hist intro}

\begin{figure}[h!]
    \centering
    \includegraphics[width=0.9\linewidth]{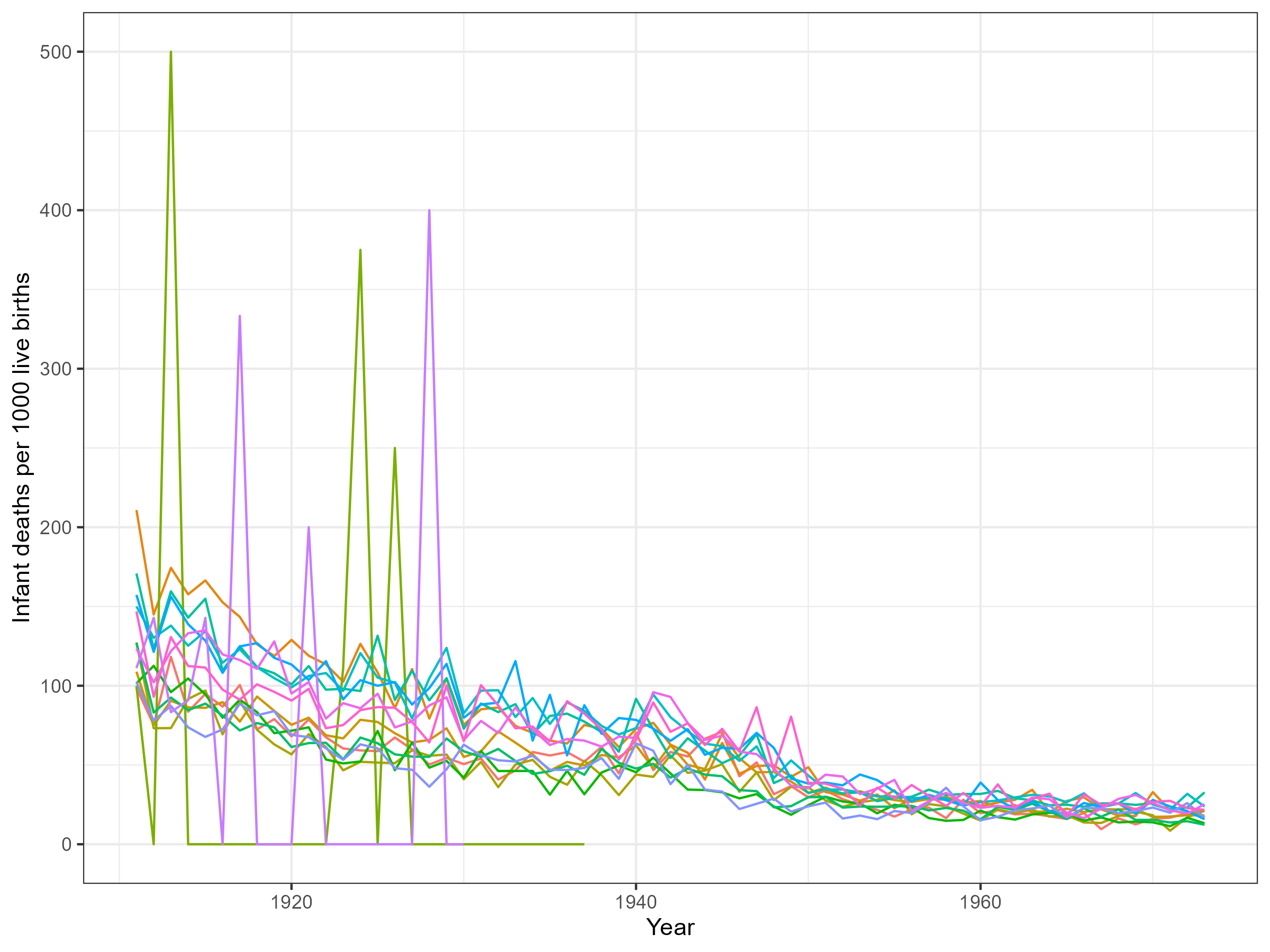}
    \caption{Infant mortality rate England and Wales, based on the raw data. Sample of $15$ local government districts depicted \citep{GBHDB_vital_statistics}: two of these are districts with populations under $500$ in $1911$ and the rest are districts with populations over $50000$.}
    \label{fig: hist IM rate raw sample Eng Wal}
\end{figure}

In this paper we consider a geographically detailed set of data collected over a key period in the history of public health in the United Kingdom: infant mortality rates by local government district in England and Wales between $1911$ and $1973$. 
During this time there is a marked drop in infant mortality rates: in $1911$ there were, on average, $130$ deaths of children under one year old per thousand births in England. This had dropped to $17$ by $1973$ \citep{GBHDB_England_Infant_Deaths, GBHDB_England_births}. There is a similar drop in Wales. 
Data are available annually for each of around 1,500 local government districts during the period
but, despite the scale and rich level of detail of the data available, how and why infant mortality declined so precipitously is not well understood. This is, in part, because the data, as it exists, is difficult to analyse: a large number of changes were made to local government district boundaries over the years of interest, meaning that --- for many districts --- we cannot compare like with like over time.  
Scholars deal with this problem by aggregating \citep{lee1991regional, woods1988causes, woods1989causes, winter1982aspects}, but these analyses miss the finer scale variations in the pattern of decline in infant mortality, such as differences between rural and urban areas. Alternatively, they have focused on fine detail but have been restricted by geographic area \citep{OUstudy} or time \citep{congdon2004small}.  
Figure \ref{fig: hist IM rate raw sample Eng Wal} shows the infant mortality rate (the number of deaths per $1000$ births) for a sample of $15$ districts with very large (over $50000$) or very small populations (less than $500$) in $1911$. 
It is clear that infant mortality declines over the period in question, but how this decline occurs varies enormously, particularly in the first $30$-$40$ years. For some local government districts, infant mortality rates begin the period very high --- over $20\%$ --- and drop steeply. For others, infant mortality rates are lower at the beginning of the period --- under $10\%$ and the decline is less marked. It is also clear that no single intervention, such as the establishment of a National Health Service in $1948$, is responsible for the decline. Furthermore, we see that as districts vary substantially in population, the variability of the series differs greatly. 

The contribution of this paper is in developing new, as well as applying
existing statistical methods, that will enable this data set to be analysed
at a fine scale over the whole of the time period in question. In particular we
show how we can automatically detect errors within time series for local authority 
districts due to missed, or inappropriate, interpolation accounting for boundary changes.
We demonstrate our method by applying it to a subset of the data. As an example of the kind of analysis that can be done with fine scale data we cluster individual series based on functional Principal Components Analysis (fPCA) --- identifying areas where infant mortality evolves in a similar fashion over the period. 
Studies that have treated mortality curves as functional data include \citet{erbas2007forecasting} and \citet{hyndman2007robust}. Those that treat them as functional data in order to cluster them include \citet{stefanucci2022analysing} 
and \citet{leger2021can}.
Clustering historical infant mortality rates according to shape has been performed on $19$th century infant mortality rates observed every $10$ years, using latent trajectory analysis rather than a functional data approach \citep{atkinson2017patterns, atkinson2017spatial}. It is clear from Figure \ref{fig: hist IM rate raw sample Eng Wal}, which depicts only a fraction of the infant mortality series, that it is 
difficult to distinguish trends at the fine scale, so being able to group the series according to common behaviour would enhance the understanding of this data set. 

Before we can apply fPCA to our data to better understand the nuances of the decline in infant mortality over the $20$th century, we first need to address the problems caused by the numerous boundary changes afflicting our data set. 
We will use interpolation to adjust those observations impacted by boundary changes.
Interpolation estimates observations for those areas and years affected, enabling
the construction of a time series covering the full period 
$1911$-$1973$ over a consistent set of boundaries. 
We use a simple interpolation approach that assumes populations are distributed evenly over area \citep{goodchild1980areal}, and introduce a process to detect situations where the interpolation is poor. 
We interpolate the raw data: annual counts of infant deaths --- those where a child dies before its first birthday --- and annual counts of births by local government district, rather than the infant mortality rate derived from these. 
This is because we want to apply our error detection process to the series of raw and interpolated counts, since it is likely that an interpolation error in the series of deaths will also occur in the series of births, then the error may be harder to detect in the proportion of infant deaths to births than in one of the series of counts individually. 
For those series in which we detect an error, as a safest approach we will merge areas that have been affected by an interpolation error, rather than trying to improve the interpolation by using a more sophisticated model (although that option is available for future analyses). 

Investigations into areal interpolation errors include \citet{liu2008accuracy}, \citet{hawley2005comparative}, \cite{gregory2002accuracy} and \citet{fisher1996modeling}. 
However, there is no agreed best method of interpolating. The performance of methods can vary according to the type of interpolation, the data available and the accuracy assessment measure. 
To the best of our knowledge the only work extant that looks to detect and assess interpolation errors --- in a data set similar to ours --- by treating them as data observed over time, as well as space, is \citet{gregory2006error}.
This work considers decennial population counts for districts, which can be approximated as normally distributed data. 
They detected changes by looking for outliers in the differenced data from one time period to the next. 
By looking at differences only over consecutive data, this method loses power by ignoring the long-term impact of a change, which will impact all future observations.
It is also not suitable for count data that does not fit the criterion for a normal approximation to Poisson data.
 
\begin{figure}[h!]
    \centering
    \includegraphics[width=0.9\linewidth]{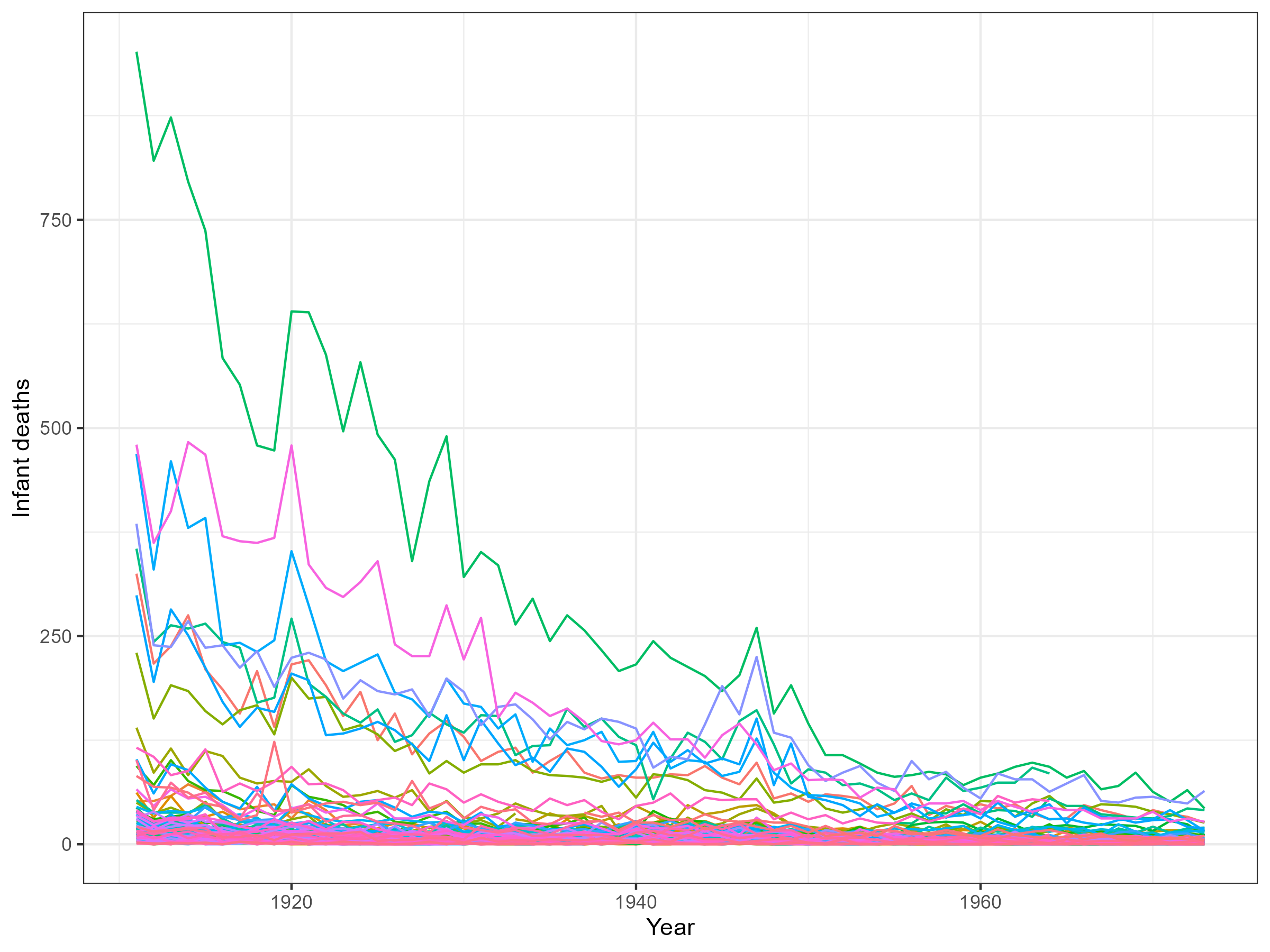}
    \caption{ Counts of infant deaths, England and Wales, based on the raw data. Random sample of $100$ local government districts depicted \citep{GBHDB_vital_statistics}.}
    \label{fig: hist I death count raw sample Eng Wal}
\end{figure}

In this paper we introduce a changepoint detection method to find the instances where interpolation has performed poorly on our data. 
An interpolation error will show as an abrupt change in the series of counts of annual infant deaths by local government district at the same time as a known  boundary change in that district.
Our data is count data with downwards trend. It includes instances of low or zero counts (Figure \ref{fig: hist I death count raw sample Eng Wal}) and is therefore not suitable for  approximation to normal.
We also directly model the trend within the data, by testing for an abrupt change in the presence of a trend, which makes it more robust than standard change-point detection methods that would ignore any trend.  
It can can be used directly on the data in our study.

Changepoint detection methods are a group of algorithms that look to identify one or more changes in a specified statistical property of a sequence, or series, of data. See, for example \citet{truong2020selective} for an overview of the extensive literature in this area. 
Our data is count data, which is most naturally modelled by a Poisson distribution. We are interested in detecting interpolation errors which would correspond to an abrupt change in the mean of the data, while allowing for gradual changes in the mean due to other factors --- such as public health improvements --- which we model as trend. We introduce a method that searches for a single abrupt change in a sequence of Poisson distributed data that is subject to trend. 
\citet{pein2021change} is among recent work to explore the detection of an abrupt change in data with a slowly evolving trend, but the method is not suitable for count data. 
Methods that search for a changepoint in Poisson distributed data, such as \cite{paparas2023maximum} and \cite{samuel1998identifying},  either ignore trend or detect changes that would be due to changes in the trend, or the introduction of trend \citep{perry2007change}.  
\cite{loader1992log} consider the detection of an abrupt change in a Poisson process with trend. This method is not suited to our data since it is developed for modelling arrival times not the number of occurrences in a sequence of time periods of equal length.

In the rest of this paper we give an overview of the data under consideration and the interpolation method applied to it (Section \ref{sec: hist the data}) ; then we introduce our changepoint detection method (Section \ref{sec: hist changepoint detection}) and give an overview of fPCA (Section \ref{sec: hist method FPCA clustering}). 
We return to the application in Section \ref{sec: hist application - CPD} where we use our changepoint detection method to identify errors in the interpolated time series of infant death counts, and, in Section \ref{sec: hist application - FPCA clustering}, we show that correcting interpolation errors affects the clustering of the infant mortality curves, and discuss some potential insights gleaned from looking at the data at this fine scale. We conclude in Section \ref{sec: hist discussion}.

\section{Data and Preliminaries}\label{sec: hist the data}

\begin{figure}[h!]
    \centering    \includegraphics[width=0.8\linewidth]{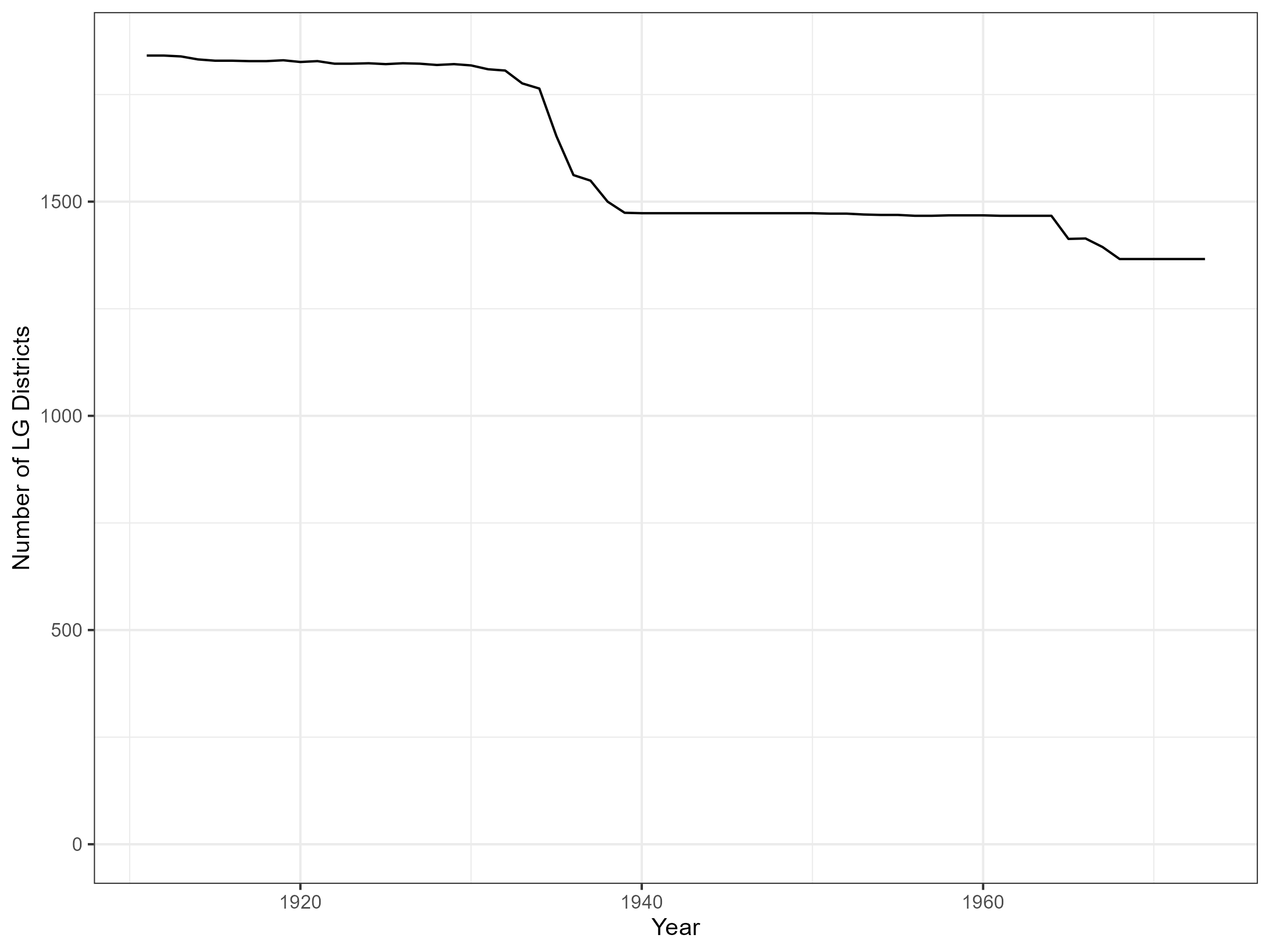}
    \caption{Counts of LGD names in England and Wales per year (raw data) \citep{GBHDB_vital_statistics}.}
    \label{fig: hist counts LGD}
\end{figure}

As mentioned in Section \ref{sec: hist intro}, the data we consider consists of annual counts of births and infant deaths recorded for local government districts in England and Wales between $1911$ and $1973$. 
These were published in the Annual Reports of the Registrar General $1911-1920$, and thereafter in the Registrar General's Statistical Review of England and Wales. These were digitised as part of the Great Britain Historical Database and are held in a vital statistics data set that is made available by the UK Data Service \citep{GBHDB_vital_statistics}.
From these data we are able to calculate the infant mortality rate per 1000 births in each district. 
However, this is not straightforward. The boundaries of the approximately $1300-1800$ geographical units for which annual data is collected (Figure \ref{fig: hist counts LGD}) undergo considerable change over the period: boundaries are adjusted, districts are created and others abolished, and names were changed. 
Boundary changes occurring between $1930$ and $1973$ are held by the GBHDB and published in their vital statistics dataset. We refer the reader to \citet{gregory2002great} and \citet{gregory2002time} for more detail on the the construction of these data collections.

The second source of data used in this study are GIS files of local government district boundaries \citep{GBHDB_LGD_bounday_files}. These are published for each census year in the period of interest: $1911$, $1921$, $1931$, $1951$, $1961$ and $1971$. Using these files and the database of annual infant mortality statistics by local government districts, we can interpolate all of the data onto the $1971$ boundaries.

While the available data covers all of England and Wales, we focus on two trial areas in this paper because of 
the large number of districts in the data set. These are the local government districts in the county of Northumberland and in East and West Sussex. We choose these areas because the county boundary of Northumberland and the combined county boundaries of East and West Sussex do not change between $1911$ and $1973$, allowing us to focus on the changes to the local government district boundaries \citep{GBHDB_E_Sussex_changes, GBHDB_W_Sussex_changes, GBHDB_Northumberland_changes}. 
We aim to create a consistent time series for all local government districts from $1911-1973$ based on the $1971$ boundaries.

The process of interpolating the data involves estimating observations for districts which have undergone boundary changes, for the years before that boundary change. For example, if a district undergoes single boundary change over the period, say, in $1935$, the observations for 
the years $1911-1934$ 
have been collected for different geographical areas to the $1971$ boundaries, while those from $1935$ to $1973$ will have been collected on the $1971$ boundaries. Therefore the observations representing the $1971$ district for the years before the change need to be estimated.

In this work, we employ one of the simplest interpolation methods --- areal weighted interpolation \citep{goodchild1980areal}, which we implement in the \texttt{R} package \texttt{areal} \citep{prener2019areal}. 
More sophisticated interpolation methods exist but these usually involve using extra data to inform the estimates (see, for example, \citet{gregory2005breaking} and \citet{hawley2005comparative} for an overview). 
The size and complexity of our data set, and the paucity of ancillary data available --- which is, additionally, often inconsistently available over our time period --- mean that these more complex methods are not straightforwardly appropriate for our data. 
We know that areal interpolation is likely to introduce some errors because of its assumption that populations are evenly spread over area --- which is not realistic for England and Wales in the $20$th century. 
This is why it is necessary to search the interpolated series for errors after the interpolation step. We give a brief overview of the  areal weighting interpolation process below. We give more details specific to the application in Appendix Section \ref{app: hist interp}.

\begin{figure}[h!]
\minipage{0.49\textwidth}
 \includegraphics[width=\linewidth]{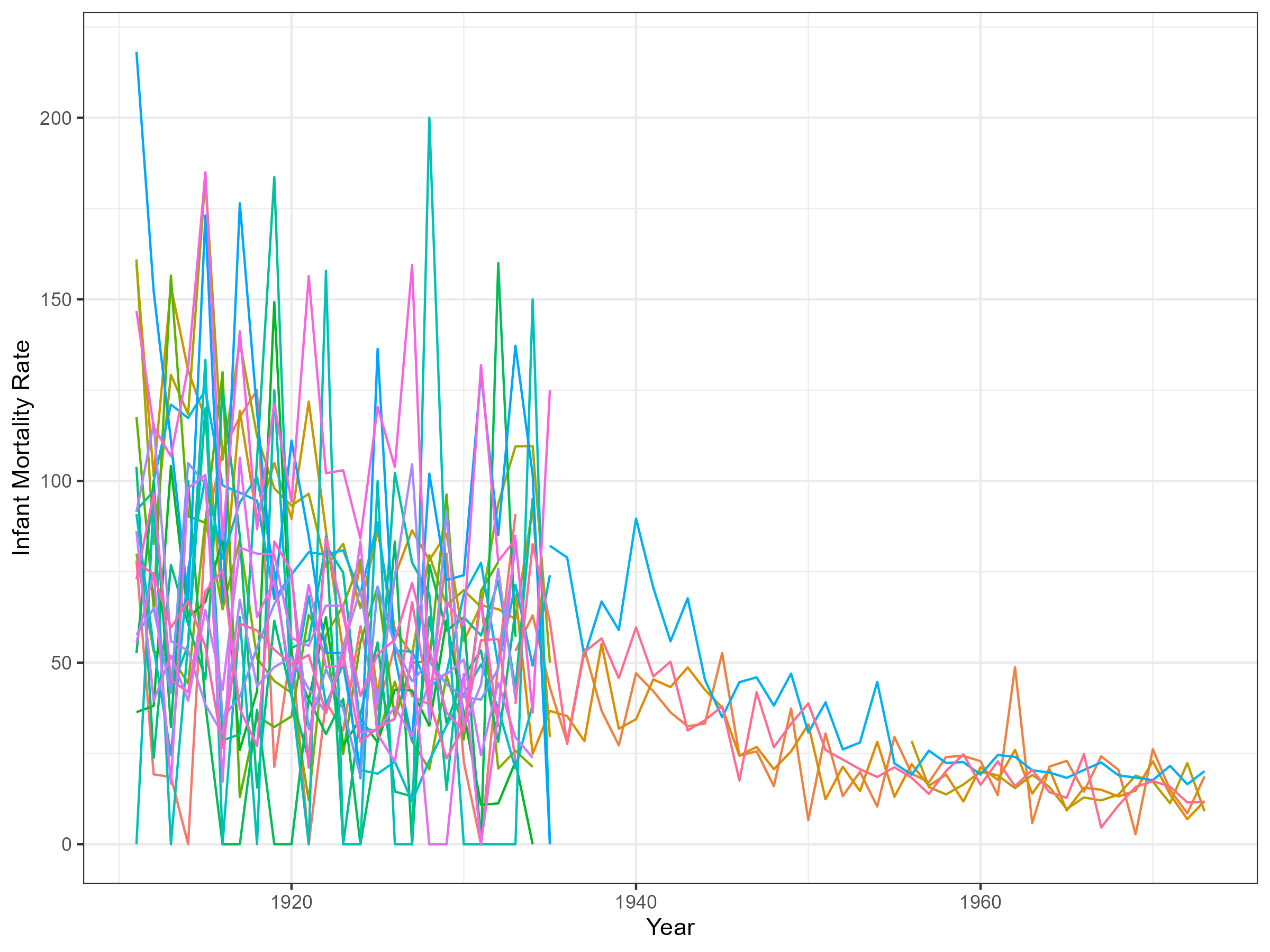}
\endminipage \hfill
\minipage{0.49\textwidth}
 \includegraphics[width=\linewidth]{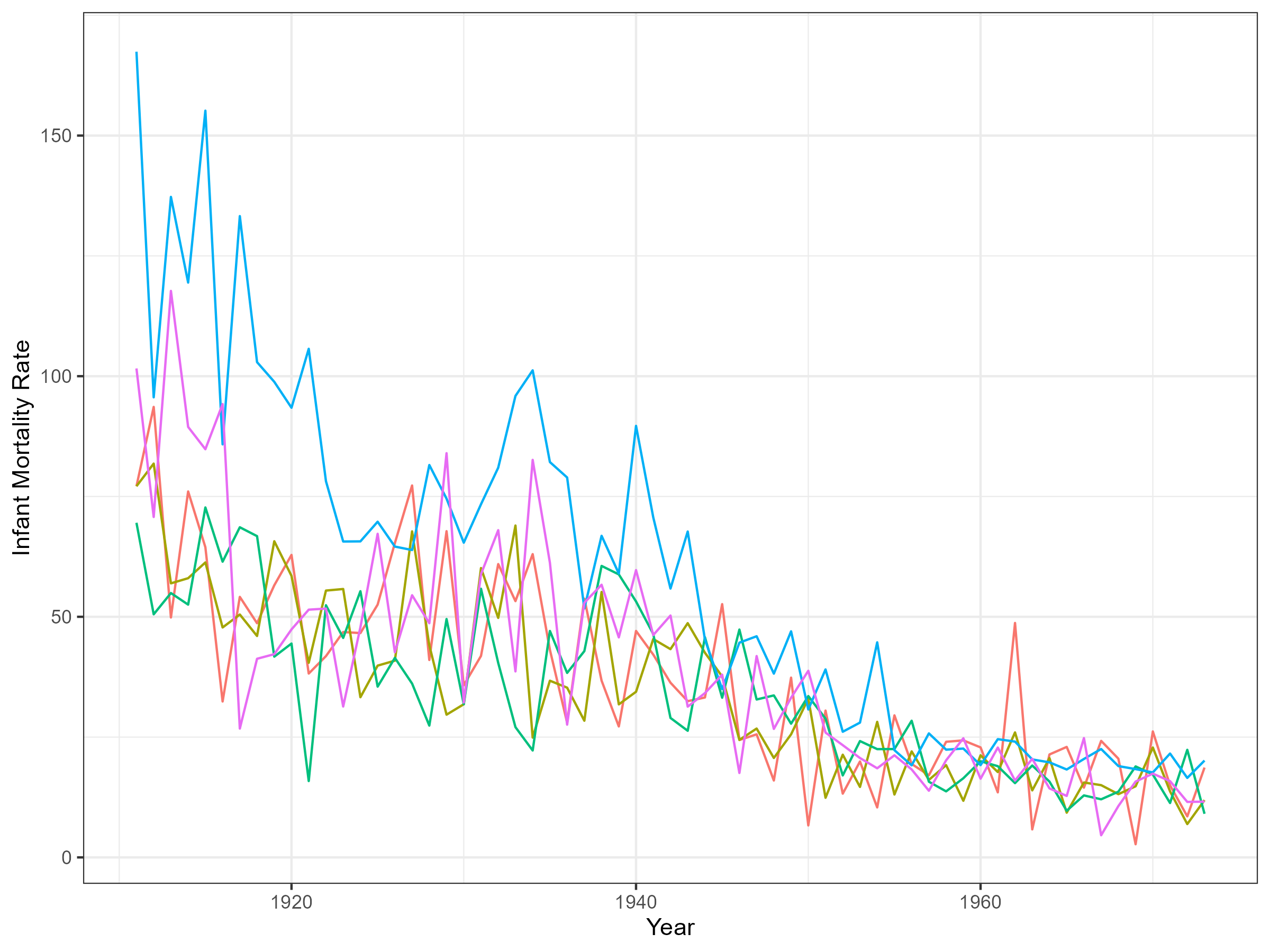}
\endminipage 
\caption{Raw and interpolated IM rates for local government districts in Sussex or Northumberland that are created or abolished $1911-1973$ \citep{GBHDB_vital_statistics}. Depicted on the left hand side is the raw data, showing large numbers of districts abolished in the mid-$1930$s and a few created that continue to have data recorded for them until $1973$. On the right hand side we see the interpolated data for $1973$ districts that were created during the period.}
\label{fig: hist im rates selected areas}
\end{figure}

We define those areas and years for which we require estimates as target districts, following the notation of \cite{gregory2006error} and adding notation for time. The estimate for the $j$th target district in the $t$th year is denoted $\hat{x}_{t,j}^{T}$
, $j=1, \ldots, d'$, where $d'$ is the number of target districts that are known to undergo at least one boundary change and 
$t=1911, \ldots, 1973$. 
For each year we also have a set of observed counts from source districts. These are the areas for which we have observations collected on known boundaries. The observed counts for the $i$th source area in the $t$th year is denoted $x_{t,i}^{S}$, $i=1, \ldots, s_t$, where $s_t$ is the number of source districts in the $t$th year under consideration. For each Target Area, one or more Source Districts will overlap with its boundaries, and we will use this subset of observations, together with knowledge of how much of the area of a Source District overlaps with the Target Area, to calculate an estimate for the Target Area.  
Let $A_j^{T}$ be the region of the $j$th target district and $|A_j^{T}|$ its area. Similarly, let $A_{t,i}^{S}$ be the region of the $i$th source district in the $t$th year and $|A_{t,i}^{S}|$ its area. Here $i=1,\ldots, s_{t,j}$ and $s_{t,j}$ is the number of source districts contributing to the $j$th target area in the $t$th year. Then if $|A_{t,i}^{S} \cap A_j^{T}|$ 
 is the area of intersection between the $i$th source and $j$th target district in year $t$,
\begin{align}
\hat{x}_{t,j}^T = \sum_{i=1}^{s_{t,j}}\frac{|A_{t,i}^{S} \cap A_j^{T}|}{|A_{t,i}^S|} x_{t,i}^S \nonumber.
\end{align}

We apply areal weighted interpolation to the raw counts of births per local government district per year, and the counts of infant deaths per local government district per year. From this we calculate the annual infant mortality rate per $1000$ live births. 
In Figure \ref{fig: hist im rates selected areas} we plot the interpolated infant mortality rates next to the raw data for areas that were created during the period, as well as those areas affected by their creation.

\section{Methods}\label{sec: Methods}
In this paper we use existing and new methods to perform our analysis. We first introduce a changepoint method developed for this application, before giving an overview of fPCA, which we use to cluster the infant mortality curves.

\subsection{Changepoint detection} \label{sec: hist changepoint detection}

\begin{figure}[h!]
\minipage{0.49\textwidth}
 \includegraphics[width=\linewidth]{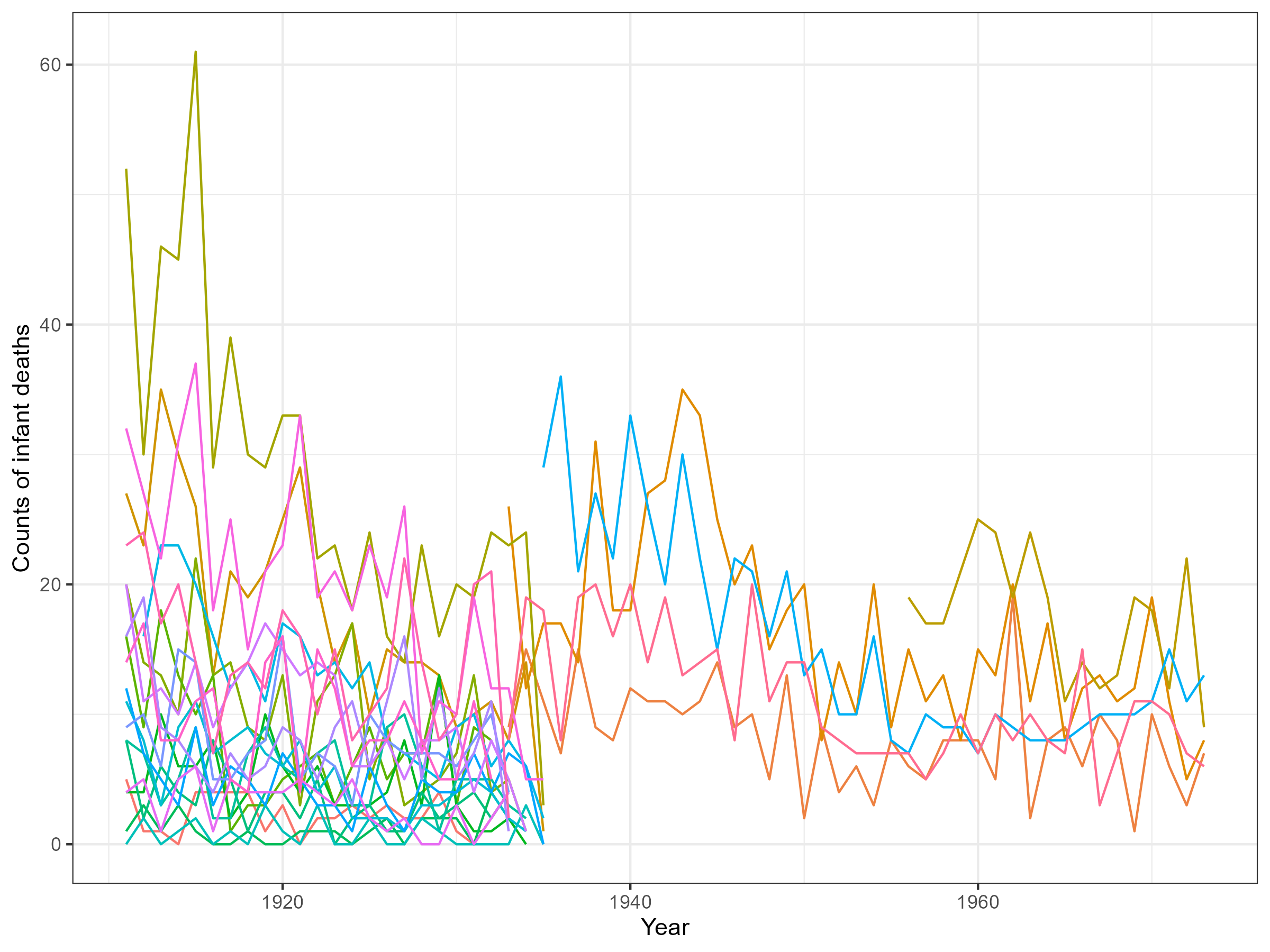}
\endminipage \hfill
\minipage{0.49\textwidth}
 \includegraphics[width=\linewidth]{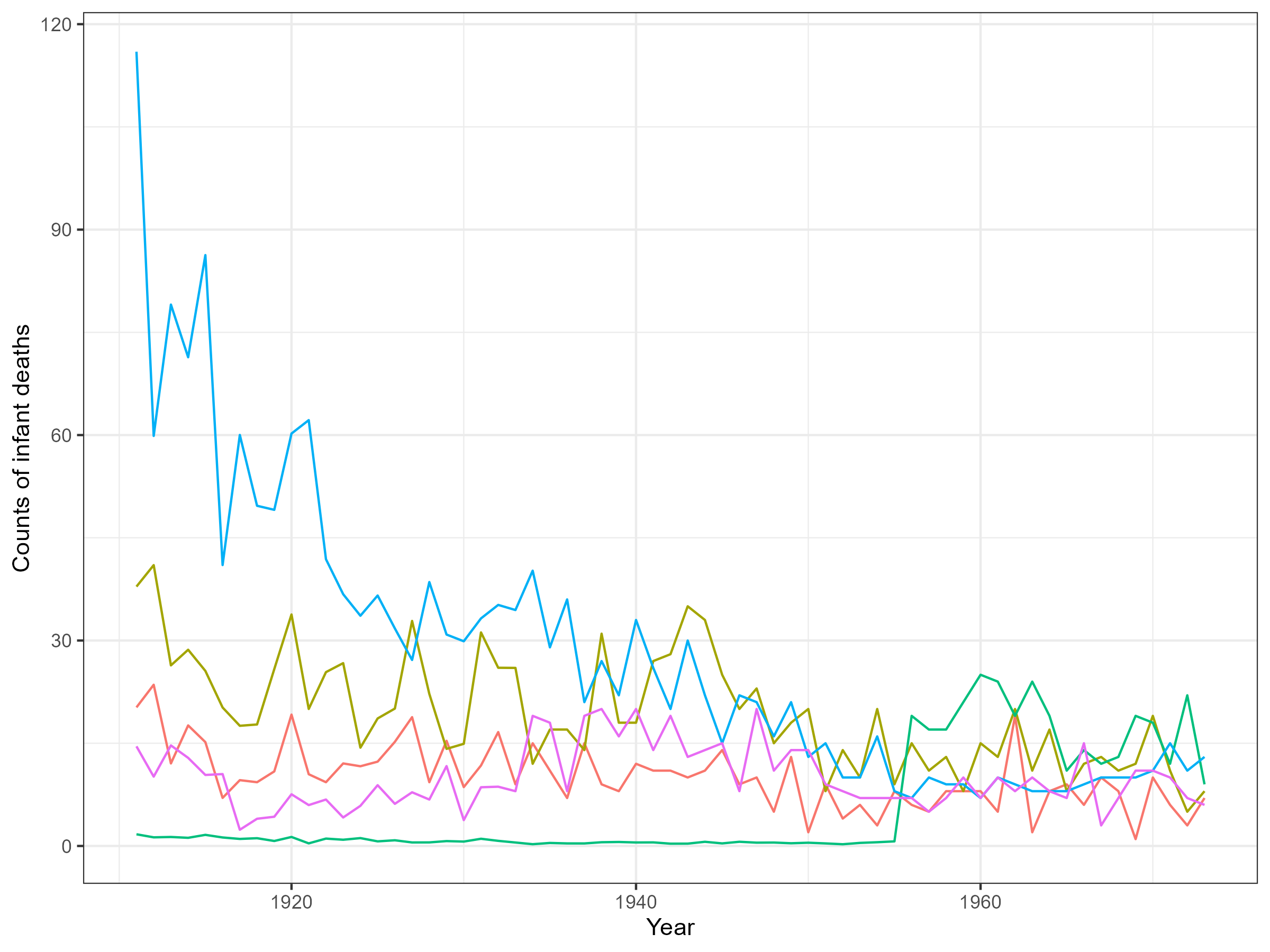}
\endminipage 
\caption{Raw and interpolated counts of infant deaths in selected areas $1911-1973$ \citep{GBHDB_vital_statistics}. The areas are selected as in Figure \ref{fig: hist im rates selected areas}. Left image: raw counts. Right image: interpolated counts.}
\label{fig: hist i death selected areas}
\end{figure}

Although the infant mortality rate is of primary interest for historical analysis, we apply a changepoint detection technique to the interpolated counts of infant deaths. These are plotted for Northumberland and Sussex in Figure \ref{fig: hist i death selected areas} alongside the uninterpolated counts. We consider the count data, rather than the infant mortality rate, because an interpolation error in the counts of infant death is likely to be present also in the counts of births, making it harder to detect an error in the infant mortality rate. We expect an interpolation error to manifest as a sharp jump upwards or downwards in the series of infant death counts at or near the year of a known boundary change.

We introduce a Poisson distribution with a time varying parameter and a time invariant parameter, where the probability of observing count $x$ in an interval $t$ is: 
\begin{align}
    P(X_t = x_t) = \frac{\lambda_t^x e^{-\lambda_t}}{x_t!}, \nonumber
\end{align}

\noindent where $\log \lambda_t = \alpha + \beta t$.

\noindent We define a single changepoint in this process at $t = \tau$ as: 
\begin{align}
  \lambda_t =
  \begin{cases} e^{(\alpha + \beta t)} ~&{\text{ if }}~t \leq \tau~,\\ 
  e^{(\alpha^*+ \beta t)} & {\text{ if }}~ t > \tau~.\end{cases} \nonumber
\end{align}

We then \noindent give a likelihood ratio based statistic, $T_\tau$, for detecting a single change at $t=\tau$ as based on the difference between the maximised log likelihood for a change versus that for no change. 

Let $\mathcal{L}_n$ be the likelihood for the model with no change (with additive constants dropped): 
\begin{align}
    \mathcal{L}_n = -\sum_{t=1}^n e^{(\alpha + \beta t)} + \sum_{t=1}^n(\alpha + \beta t) x_t  \nonumber,
\end{align}

\noindent and $\mathcal{L}_{\tau}$ be the maximised log likelihood for a change at $\tau$ (again, dropping additive constants): 
\begin{align}
       & \mathcal{L}_{\tau} = -\sum_{t=1}^{\tau} e^{(\alpha + \beta t)} + \sum_{t=1}^{\tau}(\alpha + \beta t) x_t   \nonumber \\ 
       & -\sum_{t=\tau+1}^n e^{(\alpha^* + \beta t)} + \sum_{t=\tau+1}^n(\alpha^* + \beta t) x_t  \nonumber.
\end{align}

Then $T_{\tau}$ is 
\begin{align}
   T_{\tau} = 
    2\big{[} \max_{ \alpha, \alpha^*, \beta} \mathcal{L}_\tau - \max_{\alpha, \beta} \mathcal{L}_n\big{]} \nonumber
\end{align}

\noindent This is calculated using the \texttt{glm} function in \texttt{R}. The proposed candidate for a changepoint, $\hat{\tau} = \arg\max_{\tau}T_{\tau}$. Standard practice in the changepoint detection literature is to accept a change at $\hat{\tau}$ if $T_{\tau} > c$ where $c$ is some pre-determined constant.
The thresholds for a given significance level can be calculated using simulation. However, for our use, we will require subsequent analysis for each identified change to see whether it is likely due to interpolation error after a boundary change, or is due to other historical effects. Thus we take a different approach of using the test statistic to rank regions from those with the strongest evidence of a change, so that an analyst can identify regions most important to investigate further. 

We demonstrate this method in a variety of simulated scenarios. These scenarios and accuracy measures reported over $1000$ repetitions of them are described in the Appendix, Section \ref{app: hist sim study}.

\subsection{Functional Principal Components Analysis} \label{sec: hist method FPCA clustering}

As mentioned in Section \ref{sec: hist intro},  we want to study our data at fine scale. An example of the kind of analysis that can be done using this data is clustering the data based on its functional principal components. 

To do this we treat the interpolated infant mortality curves as functional data. Functional data analysis views the underlying signal as being a continuous function of time that we are observing, with some error, at discrete time points. It uses assumptions about the continuity and smoothness of the function to help estimate it. In our case we conceive of the infant mortality curves as ones that are more or less continuously evolving, but are observed once a year only, due to the data collection method. We then use FDA to estimate the function that represents the evolving infant mortality rate for each district.

Following \citet{ramsay2005} we 
introduce a real-valued function $u_j(t)$, $j=1,\ldots, d$, which is a member of $L^2$ and is defined on the interval $\mathcal{T}$: the years $1911-1973$. 
We can consider our observations of the infant mortality rate of the $j$th district in year $t$, $y_{t,j}$ as glimpses of this continuous function (with some measurement error, denoted $\epsilon_{t,j}$): 
\begin{align}
    y_{t,j} = u_j(t) + \epsilon_{t,j} \nonumber.
\end{align}

The idea of functional PCA is that we can write the function $u_j(t)$ in terms of a set of basis functions, and view the variability in the function as due to randomness in the coefficients. Given the variability in the functions across districts, there will be specific choices of basis functions (formally defined as the eigenfunctions, $\xi$, of the covariance operator of $u(t)$) that have good properties, including the independence of the coefficients for any given district.  
\begin{align}
    \hat{u}_j(t) = \mu(t) + \sum_{p=1}^p f_{j,p} \xi_p(t), \label{eq: hist eigen basis expansion}
\end{align}

\noindent where $\xi_p$ is the eigenfunction associated with the $p$th principal component and $f_{j,p}$ is the functional principal component score for the $j$th series. The set of eigenfunctions are orthonormal.

Functional PCA estimates the covariance operator of $u(t)$ from the data and thus gets estimates of these eigenfunctions. It then estimates the coefficients for each local government district.  
In our case we proceed as follows. We smooth each series by fitting a smoothed curve to each $y_j$ --- the observed infant mortality rate for the $j$th district --- using $12$ quadratic B-splines basis functions.  These basis functions are denoted $\phi_k$, $\phi = 1, \ldots, 12$. 
By smoothing each series each $u_j(t)$ can be expressed as a known linear combination of these functions:
\begin{align}
\hat{u}_j(t) = \sum_1^{12} c_{j,k}\phi_k(t).  
\label{eq: hist basis function expansion of u}
\end{align}

\noindent Then the variance-covariance function, 
\begin{align}
\nu(s,t) = \frac{1}{d}\sum_{j=1}^d u_j(s) u_j(t),  \nonumber
\end{align}

\noindent can be estimated from \eqref{eq: hist basis function expansion of u}. The estimates of the eigenfunctions and their corresponding principal component scores in \eqref{eq: hist eigen basis expansion} can then be estimated through matrix algebra (see, for example, \citet[pp~37-58]{ramsay2005} for more details).

\begin{figure}[h!]
    \centering
    \includegraphics[width=0.9\linewidth]{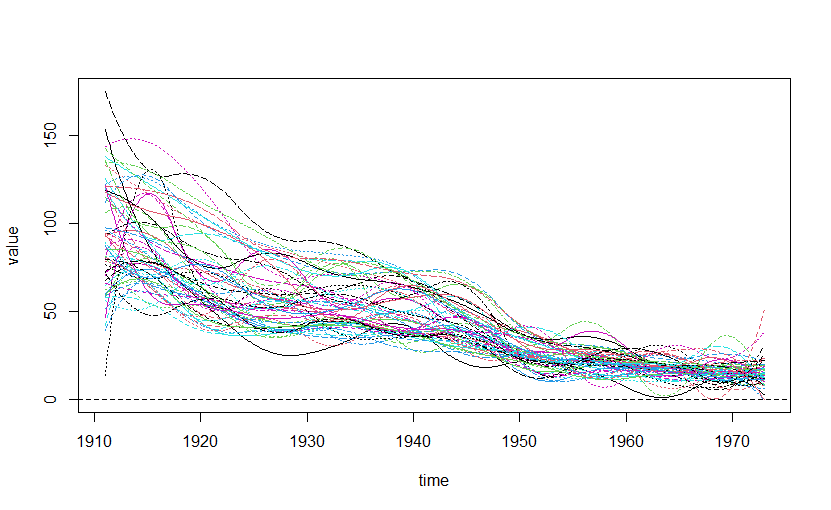}
    \caption{The smoothed series of infant mortality data. We use $12$ quadratic B-splines basis functions to smooth}
    \label{fig: hist B-spline smoothed series}
\end{figure}

The fPCA described in this paper is carried out using the \texttt{fda} package \citep{ramsay2024package} in \texttt{R}. We cluster the data based on the principal component scores, $f_{j,p}$ in \eqref{eq: hist eigen basis expansion}.
Our clustering summarises the data for each region in terms of the coefficients associated with the $p$ most important basis vectors, and clusters the data based on this summary.  So data in the same cluster will have similar values for the coefficients for each of these basis functions.
We show the first four principal eigenfunctions of the data (after adjusting for interpolation errors) in Figure \ref{fig: hist first four prin comp} in Section \ref{sec: hist application - FPCA clustering}.

We apply an agglomerative hierarchical clustering method  --- Unweighted Pair Group Method with Arithmetic Mean (UPGMA, \citet{sokal1958statistical}) using the \texttt{hclust} function in base \texttt{R}--- to the first four Functional principal components scores of the data. These account for $90\%$ of the variation about the mean curves. We determine clustering method and an optimal number of clusters by examining connectivity \citep{handl2005computational}, Dunn index \citep{dunn1974well} and Silhouette \citep{rousseeuw1987silhouettes} scores using the \texttt{clValid} package \citep{brock2008clvalid} in \texttt{R} across UPGMA and k-means \citep{hartigan1979algorithm} on data clustered after adjusting for interpolaton errors. 
We gain the top scores for UPGMA across the methods. Connectivity and Silhouette recommend two clusters, while the Dunn index has optimal values at six to nine clusters, when we cluster after correcting for data and interpolation errors. Taking this with our knowledge of the data, and that it is likely that dividing our $1500$ or so local government districts into a large number of clusters will be most helpful to historians, we select the maximum number with an optimal Dunn index score --- nine --- as the number of clusters. 
In Section \ref{sec: hist application - FPCA clustering} we discuss the fPCA functions and the clusters, including how the clustering is affected by treating areas affected interpolation errors found in \ref{sec: hist application - CPD}.

\section{Analysis}

In this section we apply the changepoint detection method described in Section \ref{sec: hist changepoint detection} to the series of infant death counts, and identify several series affected by errors. 
We cluster the data based on their fPCA scores, as described in Section \ref{sec: hist application - FPCA clustering} and show that analysing the data at a fine scale reveals interesting aspects which remain hidden when looking at this data at an aggregated scale. 
We cluster the data before and after addressing the data and interpolation errors. We show that adjusting for errors affects the clustering.

\subsection{Identifying Interpolation Errors}\label{sec: hist application - CPD}

\begin{figure}[h!]
    \centering
    \includegraphics[width=0.95\linewidth]{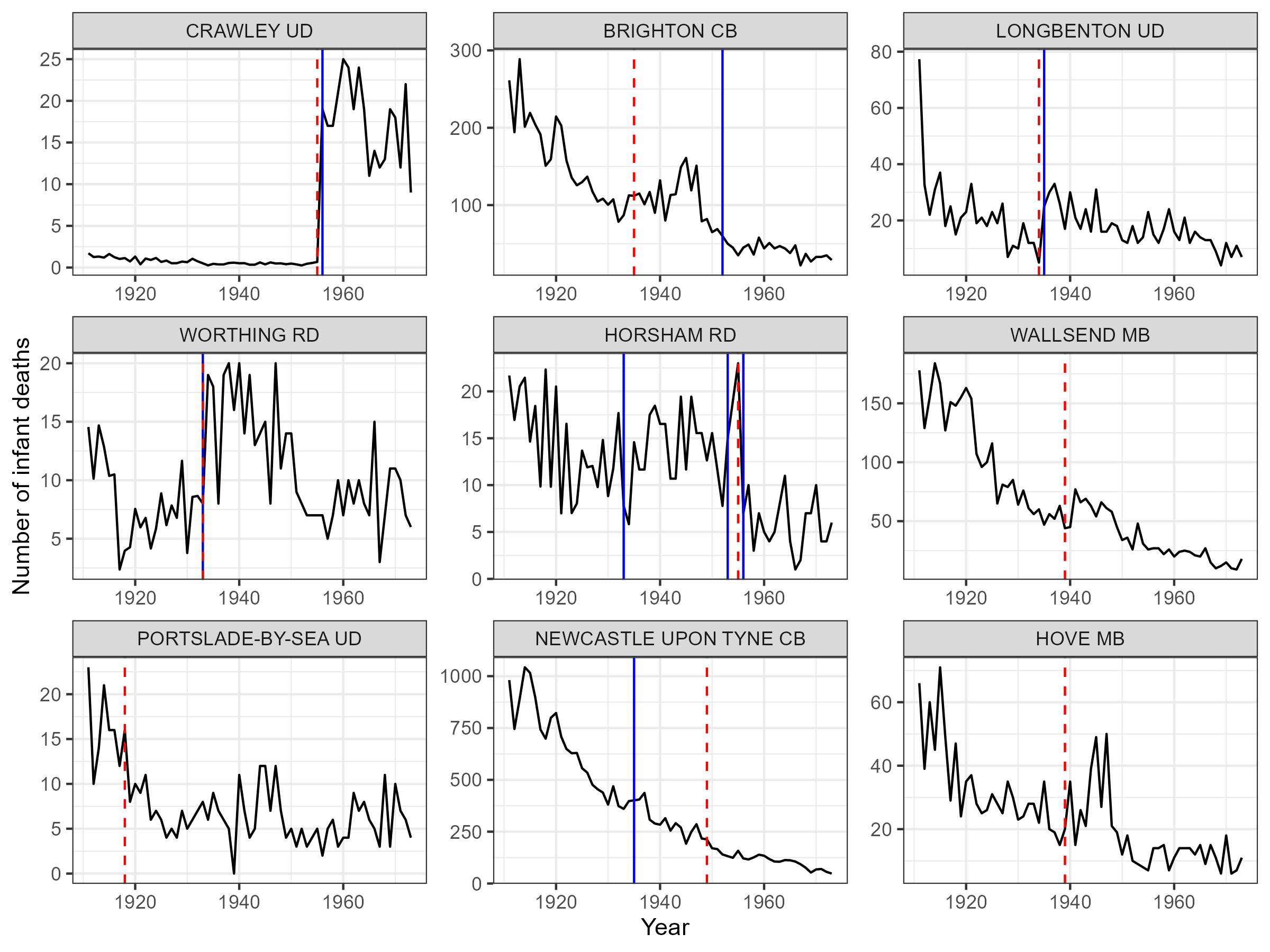}
    \caption{Results of first application of changepoint detection. Series of infant death counts with the most evidence for a changepoint, in descending order. Blue vertical line depicts the year of a known boundary change. Red dashed line depicts the date of change identified by changepoint detection.}
    \label{fig: hist top 9 changes}
\end{figure}

We apply our changepoint detection method to search each series of the interpolated counts of infant deaths for Northumberland and Sussex for a single abrupt change, while allowing for trend. As the interpolation introduces non-whole numbers, and our method is designed for count data, we round interpolated observations to the nearest whole number. 
The years $1940-1944$ are excluded from the changepoint detection analysis, given that these represent the complete years of the Second World War, which we expect to cause considerable disruption to the time series. Moreover, we know that no boundary changes were made to the local government districts in Northumberland and Sussex in that period. 
In Figure \ref{fig: hist top 9 changes} we present line plots of the interpolated series of infant mortality rates for the nine districts with the most evidence for a change in descending order of evidence. 
We also show the year of changepoints identified by our method, together with known boundary changes. As discussed in Section \ref{sec: hist intro}, we consider likely changes to be those where the identified change coincides with a known boundary change. As mentioned in Section \ref{sec: hist intro}, the identified changepoints are ranked in descending order rather than being accepted or rejected on the basis of a threshold of evidence, because the large size of the dataset and the need for potential changes to be assessed further by hand means that it is most useful for analysts to consider --- and correct --- the most egregious  errors first. 
In Appendix Figure \ref{fig: hist LR stat for top 9} we show the corresponding time series for the test statistic for a change at $\tau$ for different changepoint locations, $\tau$. These can be used to see how much uncertainty there is for the specific location of a change.  
\begin{figure}[h!]
    \centering
    \includegraphics[width=0.85\linewidth]{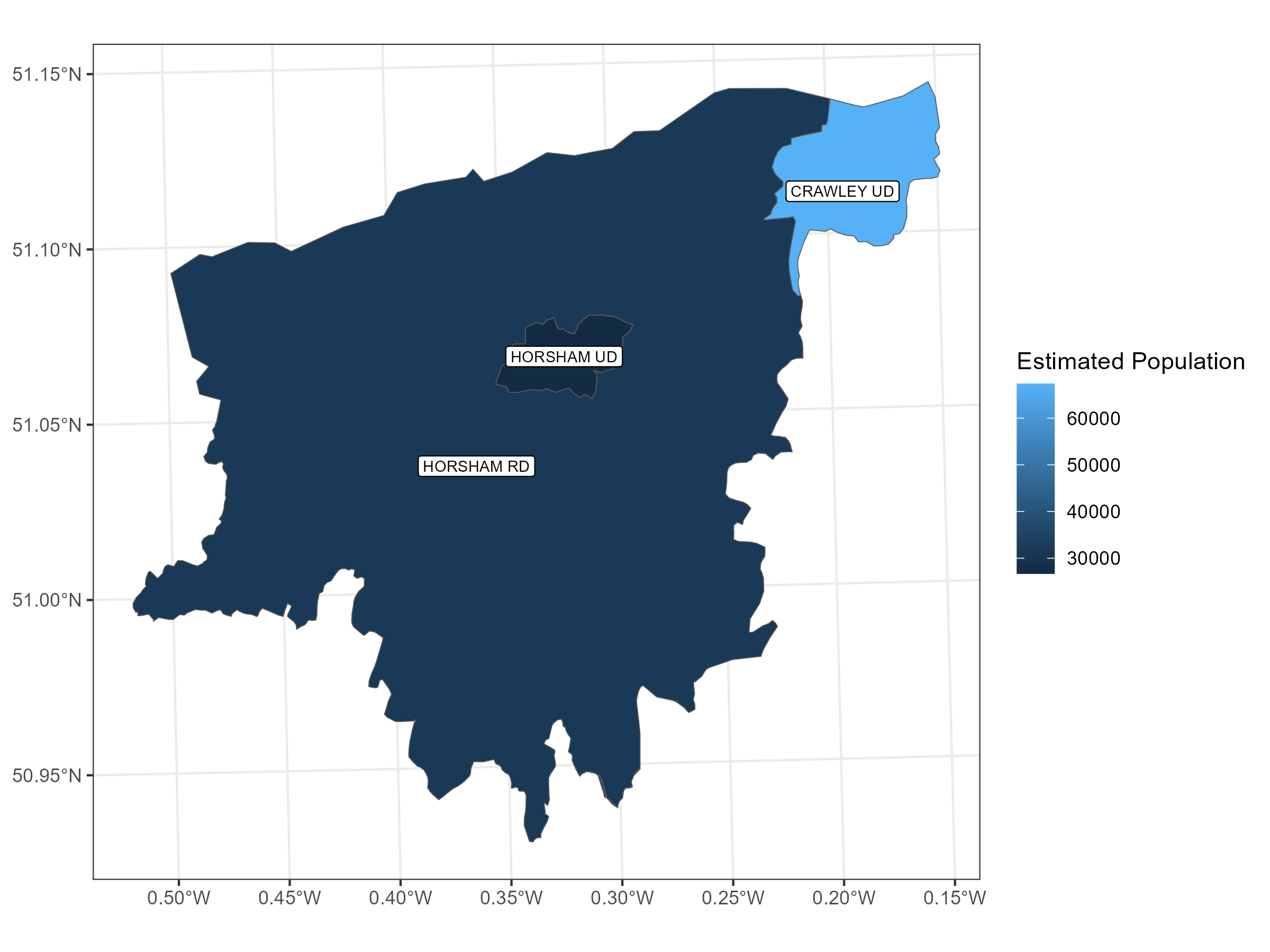}
    \caption{Horsham Rural District, Urban District and Crawley Urban District. Fill represents total estimated population in $1971$ \citep{GBHDB_vital_statistics}. Despite occupying a much smaller area than Horsham RD, Crawley UD has a considerably higher population.}
    \label{fig: hist crawley etc map plot}
\end{figure}

We give a summary of the errors detected by our changepoint detection method below. 

\begin{itemize}
    \item \textbf{Crawley Urban District (UD)}: appears to be interpolation error. Crawley UD an urban district created out of a rural district, Horsham Rural District (RD) (see Figure \ref{fig: hist crawley etc map plot} to see a map of the districts and their estimated total populations in $1971$). Interpolation method assuming that the population of Horsham is evenly distributed leads to it drastically underestimating the population occupying the area within Crawley UD's boundary before its creation. Crawley UD and Horsham RD merged. We name the adjusted district Crawley Aggregated District (AD). 

    \item \textbf{Brighton County Borough (CB) and Hove Metropolitan Borough (MB)}: there are boundary changes in $1928$ that are missing from our list of boundary changes. Brighton CB and MB Hove expand by absorbing parts of their surrounding districts \citep[pp.~691-699]{youngs1979guideI}. We adjust for this by aggregating Brighton CB with Hove MB, Newhaven RD, Steyning East RD, creating Brighton AD. 
    \item \textbf{Longbenton RD}: boundary change in $1912$ missing from our list of boundary changes. Longbenton RD was created in $1912$ \citep[p.~724]{youngs1979guideII}. We correct for this by adding the change to our interpolation process and re-interpolating.
    Longbenton RD was created in $1912$ out of part of Tynemouth RD following its abolition \citep[p.~336, pp.~347-8, p.~724, pp.~724-5]{youngs1979guideII}. 
    Without accounting for the change in $1912$ the areal interpolation does not have access to counts from source districts covered by former Tynemouth RD and pertaining to Longbenton UD from $1912$ until the known change, and relies for interpolation only upon counts from the districts that make up the expanded Longbenton UD from $1935$. The very sharp drop in count from $1911$ to $1912$ is because a count is available for Tynemouth RD in $1911$, but not for $1912$ on.   
    \item \textbf{Worthing RD}: possible interpolation error --- district created in $1935$. Corrected by merging with Chanctonbury RD, which was also created in $1933$ and from two of Worthing RD's three source districts. We call the adjusted district Worthing AD.  
\end{itemize}

Other changes, those affecting Wallsend MB, Portslade-by-Sea UD, Newcastle upon Tyne CB were investigated for potentially missed boundary changes and none were found. The identified changes are likely to be down to changes in historical trends or characteristics of the data. Notably Wallsend's identified trend is in $1939$, at the beginning of the Second World War, while Newcastle's is in the late $1940$s --- around the time the National Health Service was introduced in $1948$. Portslade-by-Sea has a low count of infant deaths through the period, so small changes in numbers can affect the curve. 

\begin{figure}[h!]
    \centering
    \includegraphics[width=0.95\linewidth]{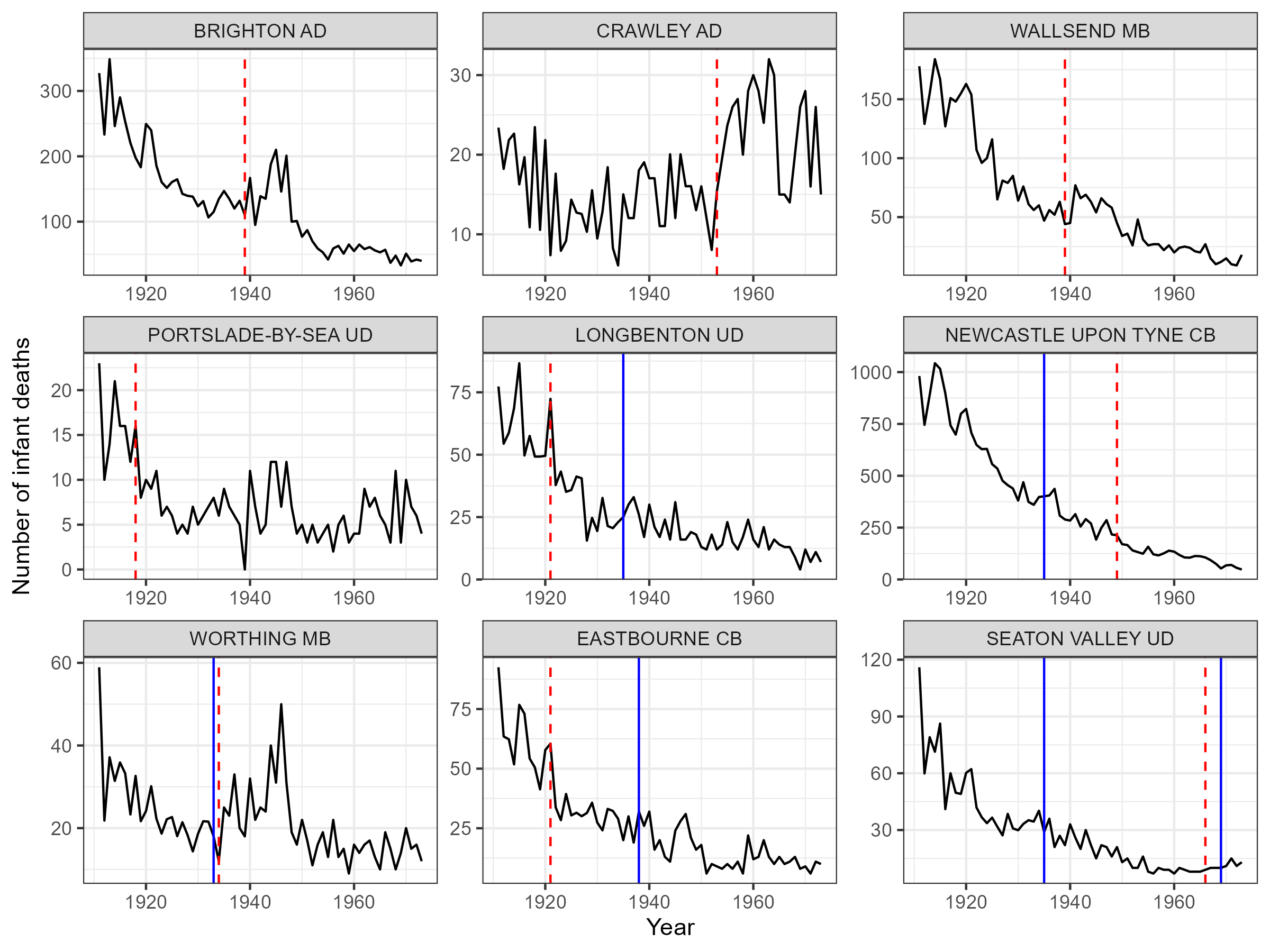}
    \caption{Results of second round of changepoint detection --- after errors have been found (Figure \ref{fig: hist top 9 changes}) and corrected. Plot depicts series of infant death counts with the most evidence for a changepoint, in descending order. As before, blue vertical line depicts the date of a known boundary change. Red dashed line depicts the date of an identified boundary change}
    \label{fig: hist top 9 changes CPD agg and corrected}
\end{figure}

Having adjusted for those errors identified as interpolation errors by amalgamating affected districts, as described in the list above, we search the series again for a changepoint. In Figure \ref{fig: hist top 9 changes CPD agg and corrected} we see that the only change identified near a known boundary change is Worthing MB. This has a smooth peak to the likelihood function (Appendix Figure \ref{fig: hist LR stat for top 9 agg and corrected}) and a low value where the function is maximised, so we decide these changes do not need adjusting for. Brighton AD has the most evidence for a change, but this changepoint is identified at the beginning of the Second World War, a point where we expect considerable disruption to population and population health, hence we accept this as a historical change.
We note that Crawley AD has strong evidence for a change in $1953$ though the likelihood test statistic has a smoother peak than in Figure \ref{fig: hist LR stat for top 9}. This changepoint is likely to be a historical change due to strong population growth in the $1950$s \citep{GBHDB_Crawley_CP_POP}. 

\subsection{Clustering the Infant Mortality Curves based on fPCA scores} \label{sec: hist application - FPCA clustering}

\begin{figure}
    \centering
    \includegraphics[width=0.99\linewidth]{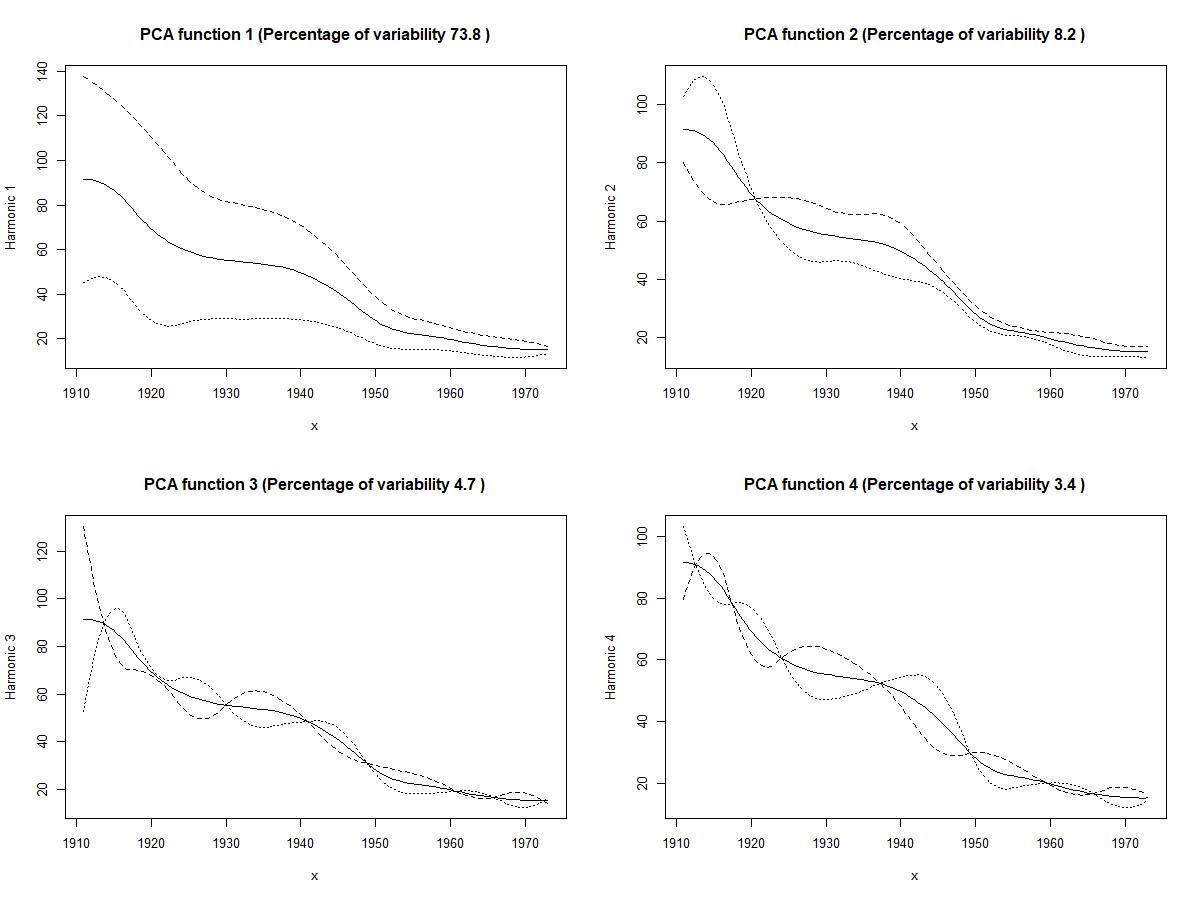}
    \caption{Plotting the first four functional principal components of the interpolated data (after correcting for error described in Section \ref{sec: hist application - CPD}. The solid line represents the mean curve of the data, while the dashed line represents a multiple of the respective principal component function added to the mean, and the dotted line a multiple of the principal component function subtracted from the mean. The multiple is set to give the effect of two standard deviations of each principal component away from the mean \citep{ramsay2005, ramsay2024package}. }
    \label{fig: hist first four prin comp}
\end{figure}

We now cluster the series of infant mortality rates using the functional PCA approach described in Section \ref{sec: hist method FPCA clustering}.
We first cluster the series before correcting for data and interpolation errors, and we compare the clustering to that after adjusting for errors. Following the  identification of errors in the interpolation, the two areas affected by the most severe errors are adjusted in a conservative way by amalgamating them with areas that are related to the same boundary change. 
In this section we first discuss the shape of the principal component functions, which give insight into the variability of the decline in infant mortality rates in Sussex and Northumberland between $1911$ and $1973$, before discussing the final clusters (formed from data after adjusting for errors). Finally we examine the way the clustering is affected by correcting for interpolation errors. 

The four fPCA functions constructed from the data that has been corrected for data and interpolation errors are depicted in Figure \ref{fig: hist first four prin comp}, plotted against the mean curve. 
The mean curve shows reasonably steady decline that begins to slow in the middle of the $1920$s. Decline is steady but slow until the mid-$1940s$, where there is another acceleration in the decline. By the $1950$s there is much less variation and the series decline slowly but steadily.
The first fPCA function, which accounts for just under three-quarters of the variability in the data, represents the varied starting points for the different districts over the period. 
The second fPCA function accounts for $8.2 \%$  of the variability. 
and shows differing behaviour of districts during the $1910$s and the $1920$s-$1930$s.
The time periods with the most variability around fPCA functions $3$ and $4$ are the first half of the $1910$s, the mid $1920$s and mid $1930$s (fPCA function $3$); and the mid $1920$s to mid $1930$s, plus the late $1930$s to $1950$ (fPCA function $4$).  

This variability shows some of the useful insight that can be gained by looking at the data on a fine scale, rather than aggregating to county level. Focusing on the the years of the First World I, for example, we see a different picture to that of  \citet{winter1982aspects}, which examines county level data and finds that for the vast majority of counties in Britain, infant mortality declines during the First World War. fPCAs $2$, $3$ and $4$ suggest the reality was more nuanced.

\begin{figure}
    \centering
    \includegraphics[width=0.9\linewidth]{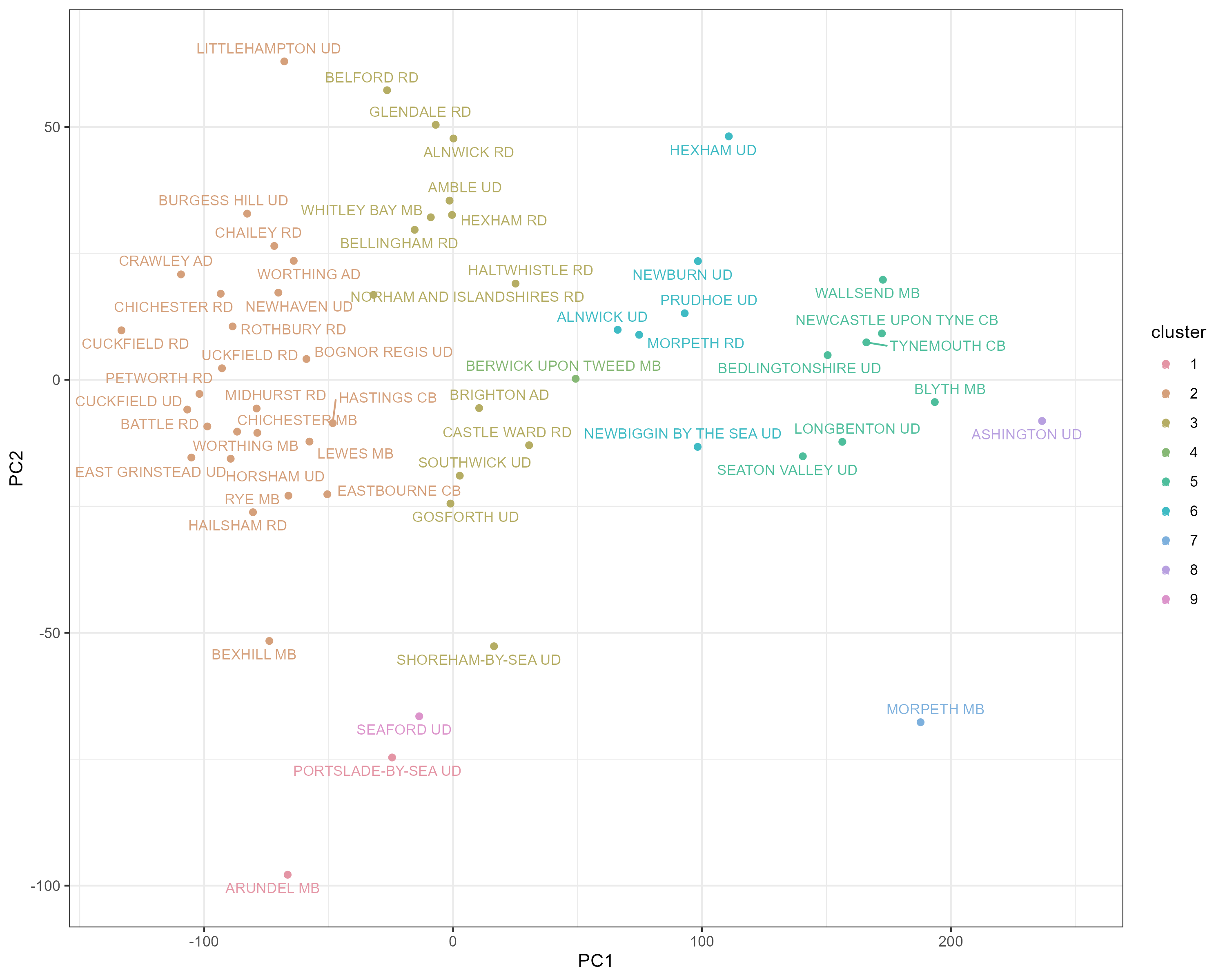}
    \caption{Scatter plot of first two functional principal component scores --- local government districts plotted by cluster, after having been adjusted for interpolation errors.}
    \label{fig: hist cluster scatter plot pc1 and 2 adjusted}
\end{figure}

\begin{figure}
    \centering
    \includegraphics[width=0.9\linewidth]{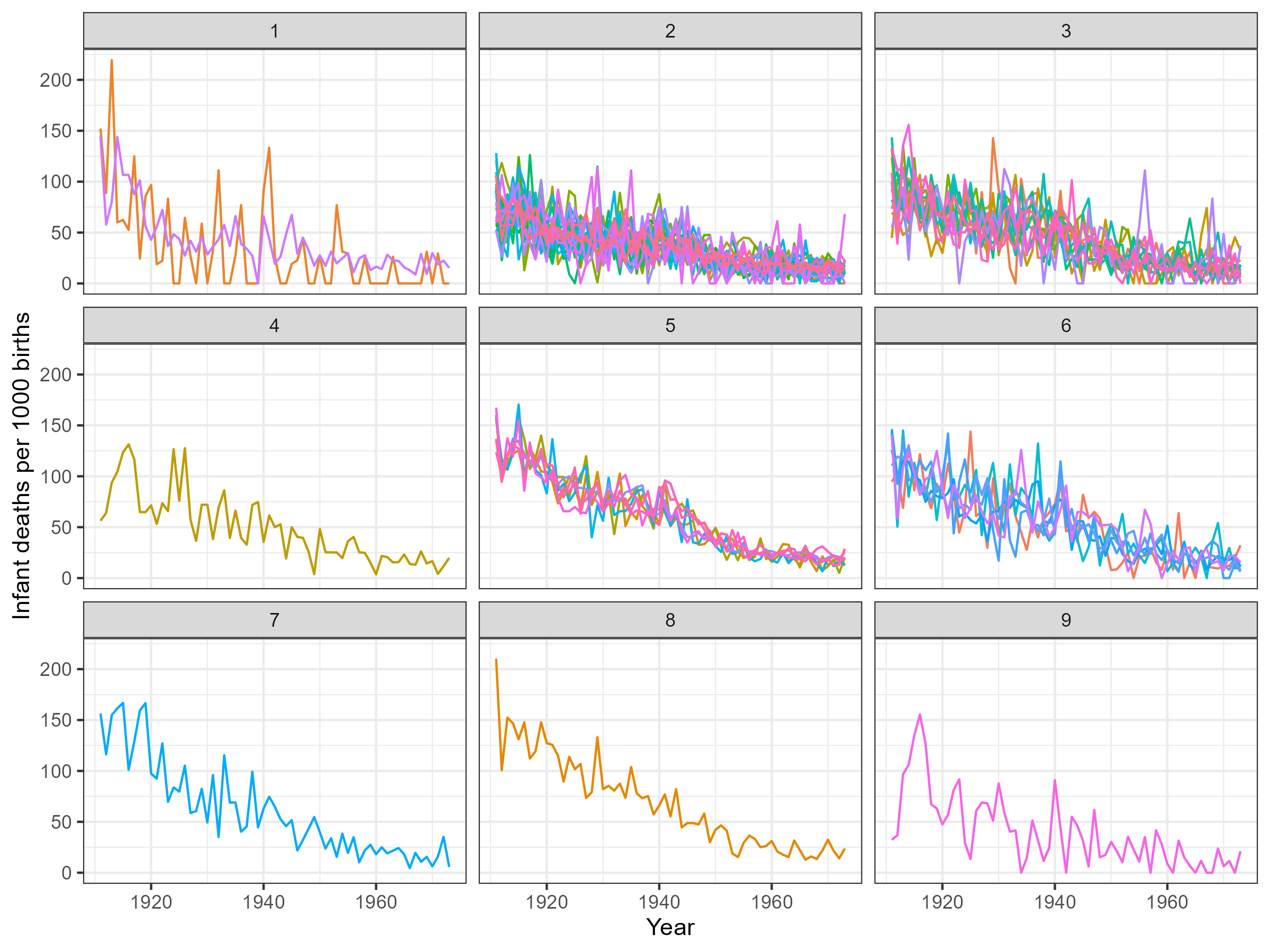}
    \caption{Local government district infant mortality rates plotted against time. Districts are grouped by cluster after adjusting for interpolation errors.}
    \label{fig: hist cluster line plot adjusted}
\end{figure}

\begin{figure}[h!]
    \centering
    \includegraphics[width=0.85\linewidth]{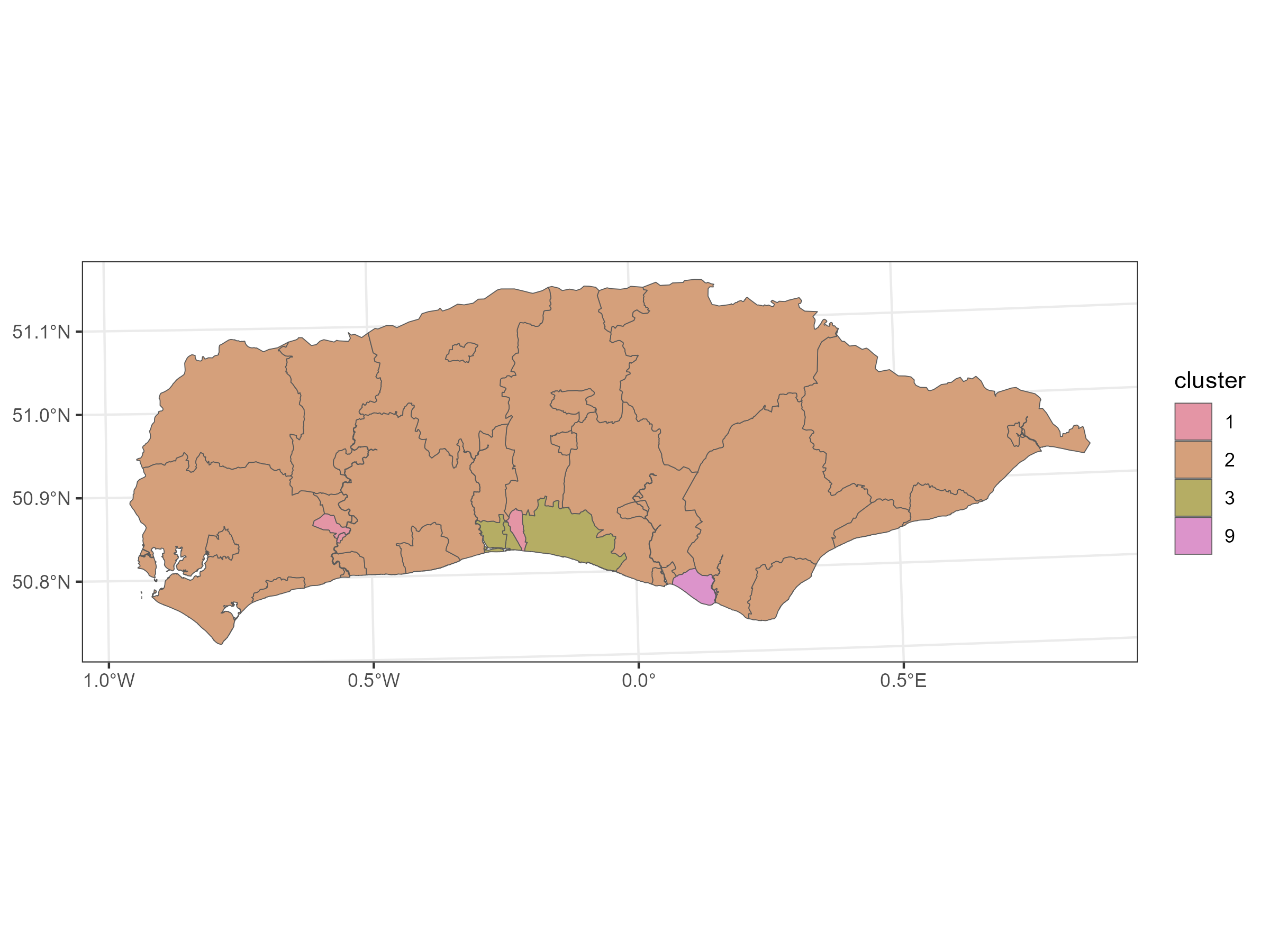}
    \caption{East and West Sussex: clusters after adjusting for interpolation errors.}
    \label{fig: hist sussex clusters map}
\end{figure}

\begin{figure}
    \centering
    \includegraphics[width=0.75\linewidth]{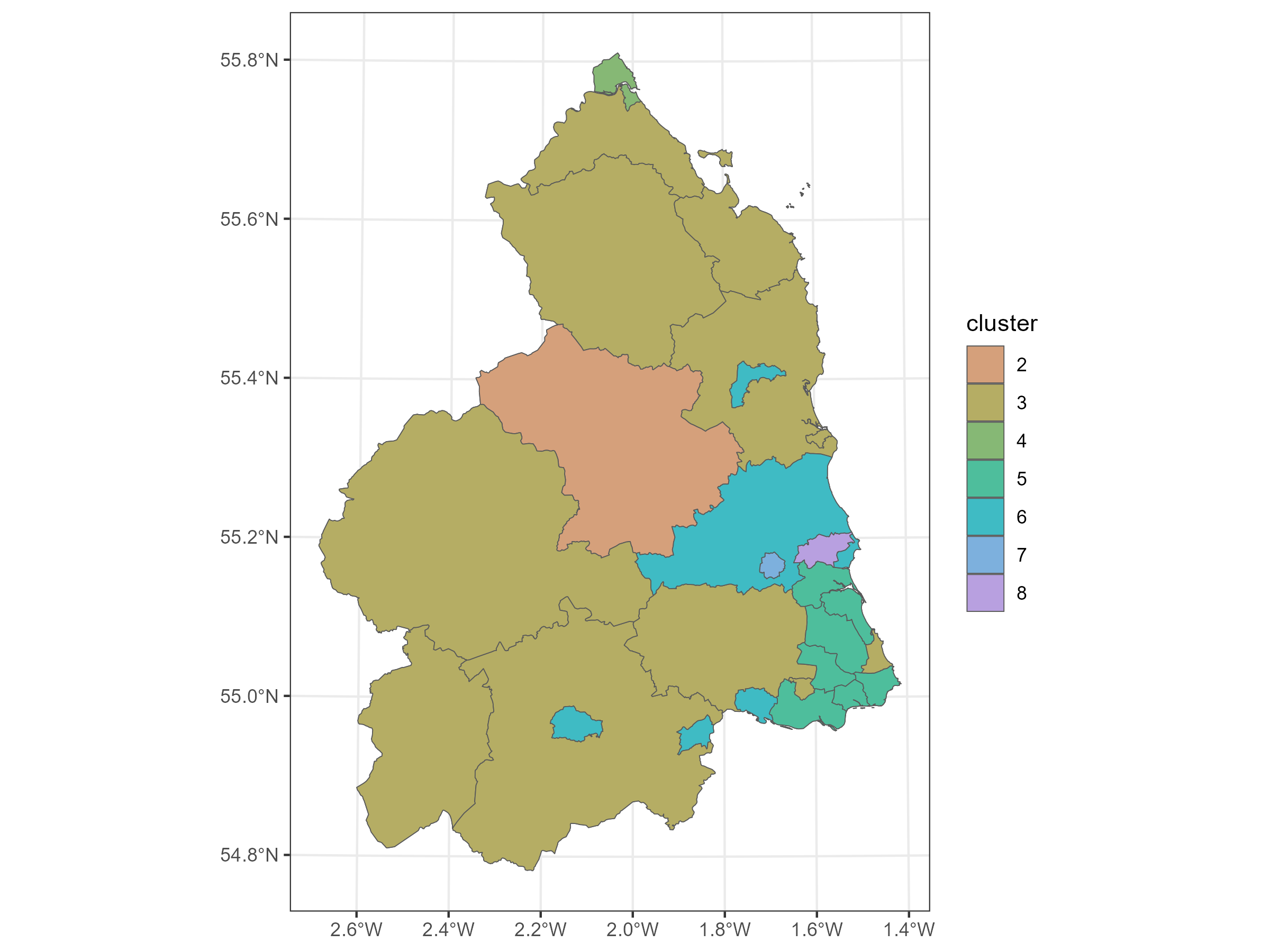}
    \caption{Northumberland: clusters after adjusting for interpolation errors.}
    \label{fig: hist northumberland clusters map}
\end{figure}

We now turn to considering the clustering of the data. 
Figure \ref{fig: hist cluster scatter plot pc1 and 2 adjusted} shows the first two principal component scores on a scatter plot, while Figure \ref{fig: hist cluster line plot adjusted} shows the infant mortality rates of local government districts over time, grouped by cluster. 
We can see some evidence of a North-South divide in the clustering (Figures \ref{fig: hist sussex clusters map} and \ref{fig: hist northumberland clusters map}) --- although it is also clear that the picture is more complex than that. 
Cluster $2$, which appears to be the healthiest cluster through the period, starting with comparatively low infant mortality rates that then decline slowly through to the $1930$s, is almost entirely local government districts in Sussex, with  Rothbury RD the only one from Northumberland in this cluster.
By contrast, clusters $8$, $7$ $6$ and $4$ --- which (in descending order) score most highly on principal component $1$ --- are exclusively made up of areas in Northumberland.

Ashington UD sits in its own cluster, cluster $8$. It has an exceptionally high infant mortality rate at the beginning of the period --- over $20\%$ of births --- and this declines in a more or less linear fashion until the late $1940$s, as opposed to the accelerated decline  over the $1910$s and early $1920$s that we see in the mean curve. The Medical Officer of Health's (MoH) report in $1911$ \citep{MOH_Ashington_1911} states that the largest cause of infant deaths was diarrhoea and enteritis, at $64$, and is attributed to an epidemic over the hot weather in summer months caused by flies to contaminate food, exacerbated by ``privy ash-pits" being close to houses. The medical officer notes that improvement work to privies and ash-pits had taken place in the previous years, and recommends domestic hygiene education. 
Morpeth MB (cluster $7$) also has very high rates at the beginning of the period, but sees some spikes in the first decade of our time period that take the IM rate to a higher level than $1911$, and it also experiences some elevated levels in the $1930$s that interrupt the general pattern of decline. The $1925$ MoH Report \citep{MOH_Morpeth_1925} attributes high death rates and infant mortality rates in the district to overcrowding and the existence of slums, but notes that education through an Infant Welfare Centre has helped to reduce the outbreaks of diarrhoea that were common. 
Both Morpeth and Ashington are reasonably populous and it is unlikely that spikes in the infant mortality rate are down to volatility that we could expect in areas where the number of births is very low.

As discussed in Section \ref{sec: hist application - CPD}, after detecting data and interpolation errors --- and correcting the data errors --- we aggregate those areas that we believe are subject to interpolation errors. This is a conservative approach to ensure we do not introduce additional errors due to, for example, modelling assumptions were we to use a more sophisticated interpolation method. 
We now turn to examining how detecting and adjusting for errors --- both in the data itself and in the interpolation --- affects the clustering. 
The principal component scores of each district --- before and after adjusting for errors --- are plotted on principal components $1$ and $2$, and $3$ and $4$, in Appendix Figures \ref{fig: hist cluster scatter plots pc1 and 2 both} and \ref{fig: hist cluster scatter plots pc3 and 4 both}. 

Most of the areas that we find are subject to interpolation errors are in cluster $5$. As discussed, this is one of the largest clusters and also the consistently healthiest over the period.  
Crawley AD is in cluster $2$, the same as its parent clusters Horsham RD and Crawley UD. This is likely because, despite the aggregation error, the infant mortality rates are low for both districts and so they both sit in the healthiest cluster. 
Brighton AD is in cluster $3$, the same as Brighton CB was before adjustments were made. Hove MB was in cluster $2$, but it had one of the highest scores of that group on principal component $1$, placing it close to cluster $3$. 
Worthing RD is in cluster $5$, but sits close to cluster $2$. 

The most obvious change in clustering after the adjustments for an unknown boundary change and aggregation of areas where there is a suspicion of an interpolation error, is that clusters $3$ and $5$ (in the initial clustering), which have similar scores on principal component $1$ but are respectively above average and below on principal component $2$, are merged into one: cluster $3$.
The other big difference is that Ashington UD, which has the highest score of any district on principal component $1$, has been separated out into its own cluster having previously been in a cluster with other districts that scored most highly on fPC $1$, while Morpeth MB, remains in its own cluster, separate from the other districts that have the highest scores on principal component $1$.

The clusters look more clearly defined in the second clustering --- for instance Haltwhistle RD looked awkward in its initial clustering of cluster $3$, being above average on PC$2$ unlike all the other members in its cluster. After adjusting for errors and re-clustering, clusters $2$, $3$, $5$ and $6$ are the main clusters and the others containing just one or two districts that are outliers on at least one of the functional Principal Components.  

\section{Discussion}\label{sec: hist discussion}
We have demonstrated a method for detecting interpolation errors in a time series of infant death counts, accommodating the fact that the data is observations of counts that exhibit trend over time. 
We have successfully identified series with interpolation errors in two trial areas of the dataset. We have shown that adjusting for the interpolation errors changes the grouping of the data. 

Extensions to the method include: modelling the data as a multivariate time series and accounting for autocorrelation and/or spatial correlation in the series; extending the model to over-dispersed count data; searching for multiple changes in each series; allowing the trend term to slowly evolve over time.

\ifOUP\else
  \section{Competing interests}
No competing interest is declared.



\section{Acknowledgments}

The infant mortality data analysed in this paper is from the Great Britain Historical Database (GBHDB), held by the UK Data Service,
\url{DOI: http://doi.org/10.5255/UKDA-SN-9035-1}, \copyright Southall, H.R., University of Portsmouth, Mooney, G., University of Portsmouth. 
The GIS files containing digital boundaries of local government districts $1911$-$1971$ are GBHDB
\url{ DOI: http://doi.org/10.5255/UKDA-SN-9321-1} \copyright Southall, H.R., University of Portsmouth, Gregory, I., Lancaster University, Aucott, P., University of Portsmouth, Burton, N., University of Portsmouth. 
These data sets are openly licensed via \url{https://creativecommons.org/licenses/by-sa/4.0/}. 
Computing in this paper is done in \texttt{R} \citep{Rproj}. Plots were produced using \texttt{ggplot2} \citep{wickham2011ggplot2}, except Figures \ref{fig: hist B-spline smoothed series} and \ref{fig: hist first four prin comp} which were produced using the \texttt{fda} package.

This paper is based on work completed while Tessa Wilkie was part of the EPSRC funded STOR-i Centre for Doctoral Training EP/S022252/1. She would also like to acknowledge the financial support of Shell Research Ltd during her PhD. She would like to thank Edward Austin for insightful discussions on the work in this paper. 
\fi

\newpage
\begin{appendices}

\renewcommand{\appendixpagename}{\textbf{Supplementary Material for \textit{Detecting interpolation errors in infant mortality counts in 20th Century England and Wales}}} 
\begin{center}  
\appendixpagename
\end{center}

\section{Interpolation between multiple boundary changes}\label{app: hist interp}
Set $1971$ boundaries as the boundaries we want to interpolate onto. There are $j$ local government districts in $1971$. These are the Target Areas. The $A_j^T$ for the $j$th local government district in $1971$. Then for each of the $j$ local government districts in $1971$. 
\begin{enumerate}
    \item Identify years of known boundary changes for the local government district. If none, no interpolation needed
    \item For $1911$ up to, but not including the year of the first boundary change, use the $1911$ boundaries for the area of each of the source districts, and the overlap between each of the source districts and the target district. Calculate $\hat{x}_{t,j}^T$ for each of the years up to (but not including) the year of the first change. 
    \item If only one known boundary change for the district, stop. Else: 
    \item For the year of the first change, up to the year of the second change, use the next available set of boundaries for the source district. For example, if there is a change in $1934$ and the next is $1955$, then use the $1951$ boundaries for the years $1934-1954$.  
    \item Continue until interpolations completed for all years prior to the last known change. For the years from the last change until the end of the period ($1973$), the un-interpolated data are used. 
\end{enumerate}

\section{Simulation Study: changepoint methods}\label{app: hist sim study}

We test our method via simulation. 
We give an overview of the simulation scenarios below, before giving the results of each. We first simulate data from a Poisson distribution with part of the rate parameter accounting for trend and the other intercept.
We then test the performance of our changepoint method on simulated data that does not fit the model described in Section \ref{sec: hist changepoint detection}. For each simulation scenario we vary the size of change and where in the series the changepoint occurs. In Scenario $4$ we also compare our method to one that searches for a change in trend or an abrupt change. 

\begin{enumerate}
    \item Abrupt change in a sequence of Poisson distributed variables. Simulate from a Poisson distribution where \begin{align}
  E(x_t|t) =
  \begin{cases} e^{(\alpha + \beta t)} ~&{\text{ if }}~t \leq \tau~,\\ 
  e^{(\alpha^* + \beta t)} & {\text{ if }}~ t > \tau~.\end{cases} \nonumber
\end{align} $n=200$ $\tau = (25, 150)$ $\alpha^* = \alpha c$, where $c = 0.5, 0.8, 1.2, 1.5$. $\alpha \sim U(2,4)$, $\beta \sim U(-0.025, 0.025)$.

\item Change in the trend of a sequence of Poisson distributed variables. Simulate from a Poisson distribution where \begin{align}
  E(x_t|t) =
  \begin{cases} e^{(\alpha+\beta t)} ~&{\text{ if }}~t \leq \tau~,\\ 
  e^{(\alpha+\beta^* t)} & {\text{ if }}~ t > \tau~.\end{cases} \nonumber
\end{align} $n=200$ $\tau = (25, 150)$ $\beta^* = \beta c$, where $c = 0.5, 0.8, 1.2, 1.5$

\item Change in a sequence of negative binomially distributed variables. Simulate from a negative binomial distribution  NB$(k,p)$: 
\begin{align}
    p(x) = \frac{\Gamma(x+k)}{\Gamma(k)x!} p^k(1-p)^x \nonumber
\end{align}

with mean $\mu = \frac{k(1-p)}{p}$ where $k$ is the size parameter. 
This is a heavier-tailed model for count data than the Poisson. 
We simulate our negative binomial variables with a time-varying mean, $\mu_t$, and with a change at $\tau$ as follows: 
\begin{align}
  \mu_t =
  \begin{cases} e^{(\alpha + \beta_t))} ~&{\text{ if }}~t \leq \tau~,\\ 
  e^{(\alpha^*+ \beta_t))} & {\text{ if }}~ t > \tau~\end{cases} \nonumber. 
  \end{align}  

As before, $n=200$ $\tau = (25, 150)$ $\alpha^* = \alpha c$, where $c = 0.5, 0.8, 1.2, 1.5$. $\alpha \sim U(2,4)$, $\beta \sim U(-0.015, 0.015)$. $k \sim U(1,10)$ We reduce the range for $\beta$, compared with the Poisson-based simulations. This is to control the variance, which increases with $t$: $\mu_t + \frac{\mu_t^2}{k}$

    \item Abrupt change in a sequence of Poisson distributed variables without trend. Simulate from a Poisson distribution where \begin{align}
  E(x_t|t) =
  \begin{cases} e^{(\alpha)} ~&{\text{ if }}~t \leq \tau~,\\ 
  e^{(\alpha^*)} & {\text{ if }}~ t > \tau~.\end{cases} \nonumber
\end{align} $n=200$ $\tau = (25, 150)$ $\alpha^* = \alpha c$, where $c = 0.5, 0.8, 1.2, 1.5$. $\alpha \sim U(2,4)$. 

\end{enumerate}

\subsection{Results}



\begin{table}[h!]
\begin{center}
\begin{minipage}{0.65\textwidth}
\caption{Results of simulation Scenario 1 over $1000$ repetitions. Accuracy presents the percentage of repetitions where the method identifies a changepoint within $+/-5$ of the true change. True positive rate represents the number of repetitions where is changepoint is identified, regardless of accuracy. Test statistic threshold is calculated to provide a $5\%$ false positive rate, based on $1000$ repetitions with no change.} \label{tab: hist res scenario 1}
\begin{tabular}{rrrrr}
\toprule
  & \multicolumn{2}{c}{Accuracy} & \multicolumn{2}{c}{True positive rate} \\
Change Factor & $\tau = 25$ & $\tau = 150$ & $\tau = 25$  &  $\tau = 150$\\ 
  \midrule
0.5 & 99.6 & 93.6 & 100.0 & 96.6 \\ 
  0.8 & 92.2 & 78.2 & 95.0 & 85.2 \\ 
  1.2 & 96.7 & 84.4 & 98.5 & 89.0 \\ 
  1.5 & 100.0 & 98.5 & 100.0 & 99.6 \\ 
   \botrule
\end{tabular} 
\end{minipage}
\end{center}
\end{table}


\begin{table}[h!]
\begin{center}
\begin{minipage}{0.65\textwidth}    
\caption{Results of simulation Scenario $2$. Accuracy presents the percentage of repetitions where the method identifies a changepoint within $+/-5$ of the true change. True positive rate represents the number of repetitions where is changepoint is identified, regardless of accuracy. Test statistic threshold is calculated to provide a $5\%$ false positive rate, based on $1000$ repetitions with no change.} \label{tab: hist res scenario 2}
\begin{tabular}{rrrrr}
  \toprule
  & \multicolumn{2}{c}{Accuracy} & \multicolumn{2}{c}{True positive rate} \\
Change Factor & $\tau = 25$ & $\tau = 150$ & $\tau = 25$  &  $\tau = 150$\\ 
  \midrule
0.5 & 12.9 & 88.3 & 23.0 & 93.5 \\ 
  0.8 & 2.0 & 60.8 & 8.5 & 75.4 \\ 
  1.2 & 2.5 & 53.6 & 7.9 & 68.0 \\ 
  1.5 & 15.9 & 83.4 & 25.5 & 92.1 \\ 
   \botrule
\end{tabular}
\end{minipage}
\end{center}
\end{table} 

\begin{table}[h!]
\begin{center}
\begin{minipage}{0.65\textwidth} 
\caption{Results of simulation Scenario $3$. Accuracy presents the percentage of repetitions where the method identifies a changepoint within $+/-5$ of the true change. True positive rate represents the number of repetitions where is changepoint is identified, regardless of accuracy. Test statistic threshold is calculated to provide a $5\%$ false positive rate, based on $1000$ repetitions with no change.} \label{tab: hist res scenario 3}
\begin{tabular}{rrrrr}
  \toprule
  & \multicolumn{2}{c}{Accuracy} & \multicolumn{2}{c}{True positive rate} \\

Change Factor & $\tau = 25$ & $\tau = 150$ & $\tau = 25$  &  $\tau = 150$\\ 
  \midrule
0.5 & 0.0 & 25.0 & 0.2 & 25.2 \\ 
  0.8 & 0.0 & 12.9 & 1.8 & 15.0 \\ 
  1.2 & 0.0 & 10.8 & 10.2 & 17.3 \\ 
  1.5 & 0.0 & 29.0 & 16.5 & 31.7 \\ 
   \botrule
\end{tabular}
\end{minipage}
\end{center}
\end{table}

\begin{table}[h!]
\begin{center}
\begin{minipage}{0.75\textwidth} 
\caption{Results of simulation Scenario 4 over $1000$ repetitions. Method $1$ searches for an abrupt change, while Method $2$ searches for an abrupt change or a change in trend. As before, accuracy presents the percentage of repetitions where the method identifies a changepoint within $+/-5$ of the true change. True positive rate represents the number of repetitions where is changepoint is identified, regardless of accuracy. Test statistic threshold is calculated for each method to provide a $5\%$ false positive rate, based on $1000$ repetitions with no change.} \label{tab: hist res scenario 4}
\begin{tabular}{lrrrrr}
  \toprule
    & & \multicolumn{2}{c}{Accuracy} & \multicolumn{2}{c}{True positive rate} \\
Method & Change Factor & $\tau = 25$ & $\tau = 150$ & $\tau = 25$  &  $\tau = 150$ \\ 
  \midrule
   1 & 0.5 & 100.0 & 99.9 & 100.0 & 100.0 \\ 
   2 & 0.5 & 99.3 & 99.9 & 100.0 & 100.0 \\ 
   1 & 0.8 & 93.8 & 95.9 & 98.8 & 99.9 \\ 
   2 & 0.8 & 88.3 & 91.0 & 98.3 & 99.4 \\ 
   1 & 1.2 & 98.2 & 97.1 & 100.0 & 99.6 \\ 
   2 & 1.2 & 94.4 & 96.4 & 99.6 & 99.9 \\ 
   1 & 1.5 & 100.0 & 100.0 & 100.0 & 100.0 \\ 
   2 & 1.5 & 100.0 & 100.0 & 100.0 & 100.0 \\ 
   \botrule
\end{tabular} 
\end{minipage}
\end{center}
\end{table}

We see in Table \ref{tab: hist res scenario 1} that the method has high power and decent accuracy for most changepoint locations and  sizes of change in Scenario $1$. This is as we might expect, as in this scenario the method is looking to detect an abrupt change in a sequence of Poisson variables with trend, which is what it has been developed to do. We note that the repetitions where the method is less successful --- with a changepoint at $150$ and a small change --- are where $\beta$ is negative, and the process is very close to zero by $t=150$, making a change very difficult to detect. 
In Table \ref{tab: hist res scenario 4} we show the results of Scenario $4$, where we set two methods to search for an abrupt change in a series of Poisson variables without trend. We compare our method, which assumes constant trend, to one that searches for a change in trend as well as an abrupt change. We see, as we might expect, that the first method is more accurate and has more power than the second, in almost all instances --- except those where the change is easy to detect and there is little to distinguish between the performance of both methods.

In Scenario $2$ (Table \ref{tab: hist res scenario 2})  we see that the location and the size of the change have a marked impact on power and accuracy. The method has slightly more power and is more accurate when $\tau=150$ versus when $\tau = 25$. The false positives appear to be concentrated between $t=150$ and $t=n$, and appear to be where $\beta$ is positive and reasonably close to its limit of $0.025$. This means the simulated data follows an exponentially increasing pattern, and towards the end of the sequence this behaviour is being mis-identified as a shift in the time invariant parameter.
In Scenario $3$ (Table \ref{tab: hist res scenario 3}), where we have a change in variables that are not Poisson distributed, the method performs poorly for all sizes of change and change locations. The method performs very poorly when $\tau=25$, with very low power and low levels of accuracy. It is markedly more accurate when $\tau = 150$ and has slightly better power.  
It would appear that a large portion of inaccurately identified changepoints occur when the trend parameter is positive and close to its limits, which leads to a false positive identified close to $t=n$. Additionally, this inflates the penalty parameter estimated to control the false positive rate. We note that the false positive rate --- the percentage of repetitions with no change for which a change is reported --- is $86.6\%$ for Scenario $3$
when using a threshold calculated for Poisson variables (from Scenario $1$). This is due to the Negative Binomial distribution being heavier tailed than the Poisson, and is an additional reason for ranking our changepoints rather than accepting or rejecting based on a threshold, as it allows us to ignore errors in thresholds due to model error. 

\pagebreak
\section{Application: likelihood ratio test statistic plots}

\begin{figure}[h!]
    \centering
    \includegraphics[width=0.7\linewidth]{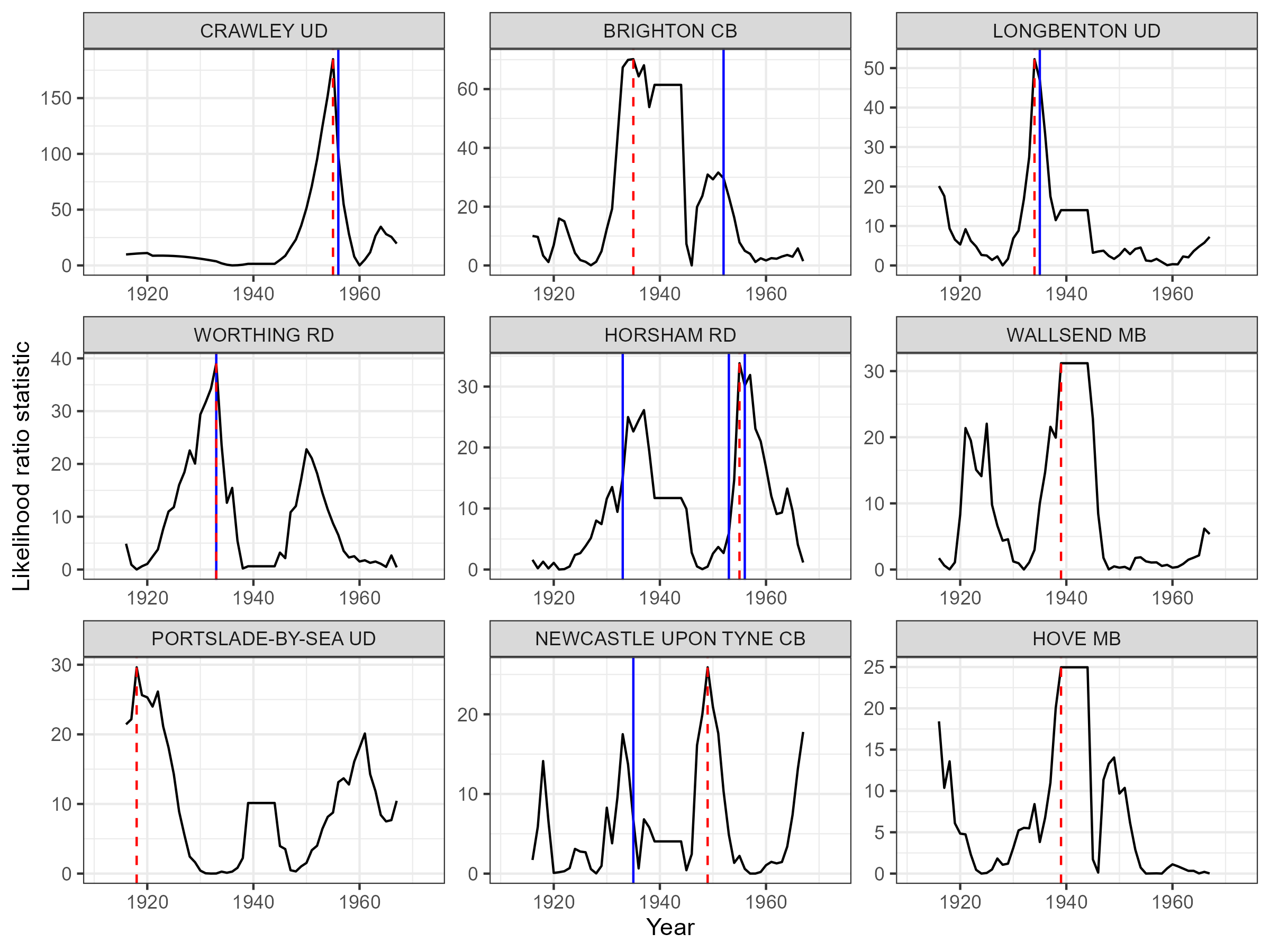}
    \caption{Likelihood ratio test statistic for a change, plotted for the top nine identified changes. Known boundary changes, and those identified by our changepoint detection method, are depicted as in Figure \ref{fig: hist top 9 changes}.}
    \label{fig: hist LR stat for top 9}
\end{figure}

\begin{figure}[h!]
    \centering
    \includegraphics[width=0.7\linewidth]{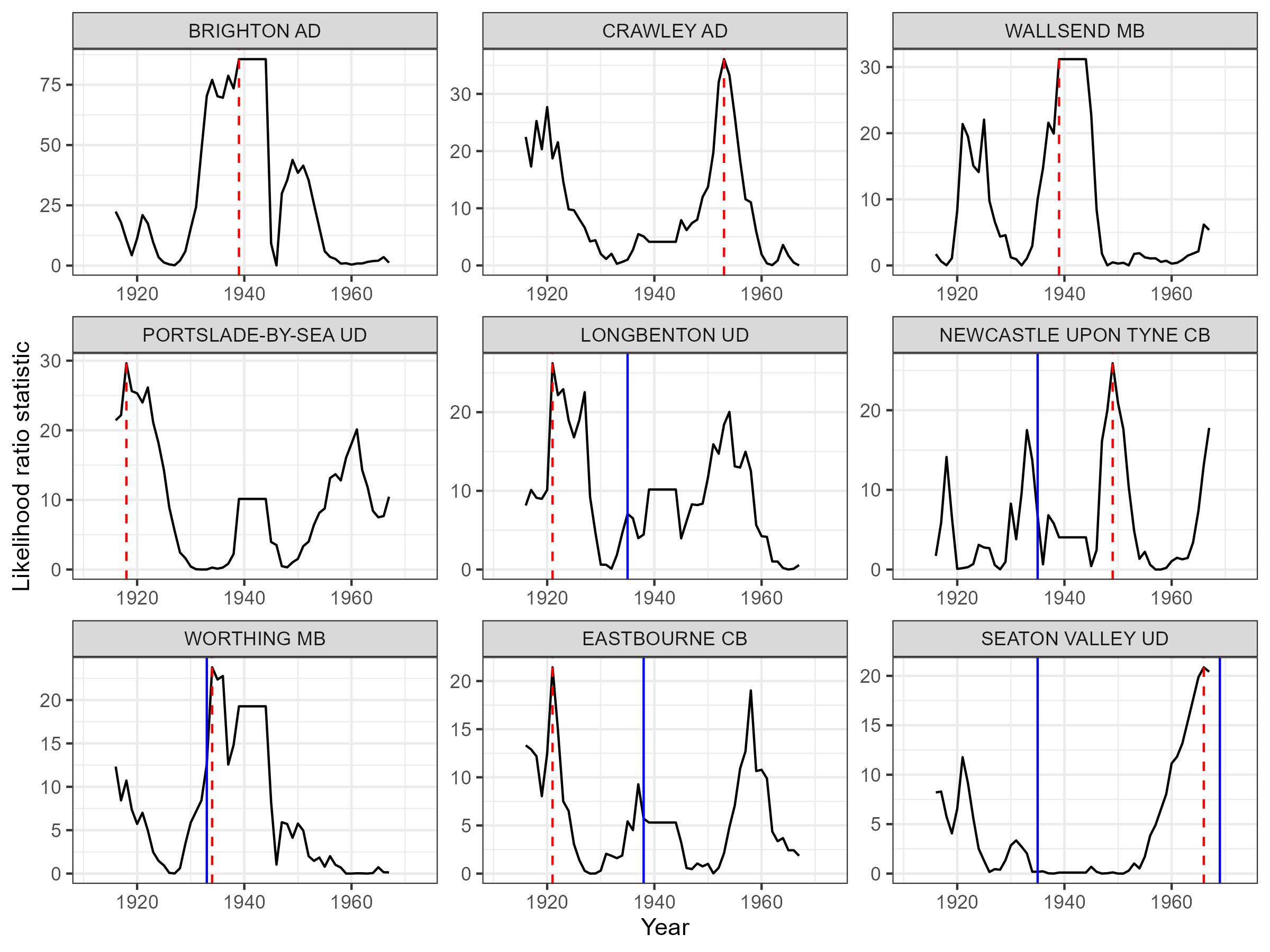}
    \caption{Likelihood ratio test statistic for a change, plotted for the top nine identified changes. Known boundary changes, and those identified by our changepoint detection method, are depicted as in Figure \ref{fig: hist top 9 changes CPD agg and corrected}.}
    \label{fig: hist LR stat for top 9 agg and corrected}
\end{figure}

\pagebreak

\section{Additional plots of fPCA clusters} \label{app: hist plots}

\begin{figure}[h!]
\centering
\begin{minipage}{0.49\linewidth}
     \includegraphics[width=0.99\linewidth]{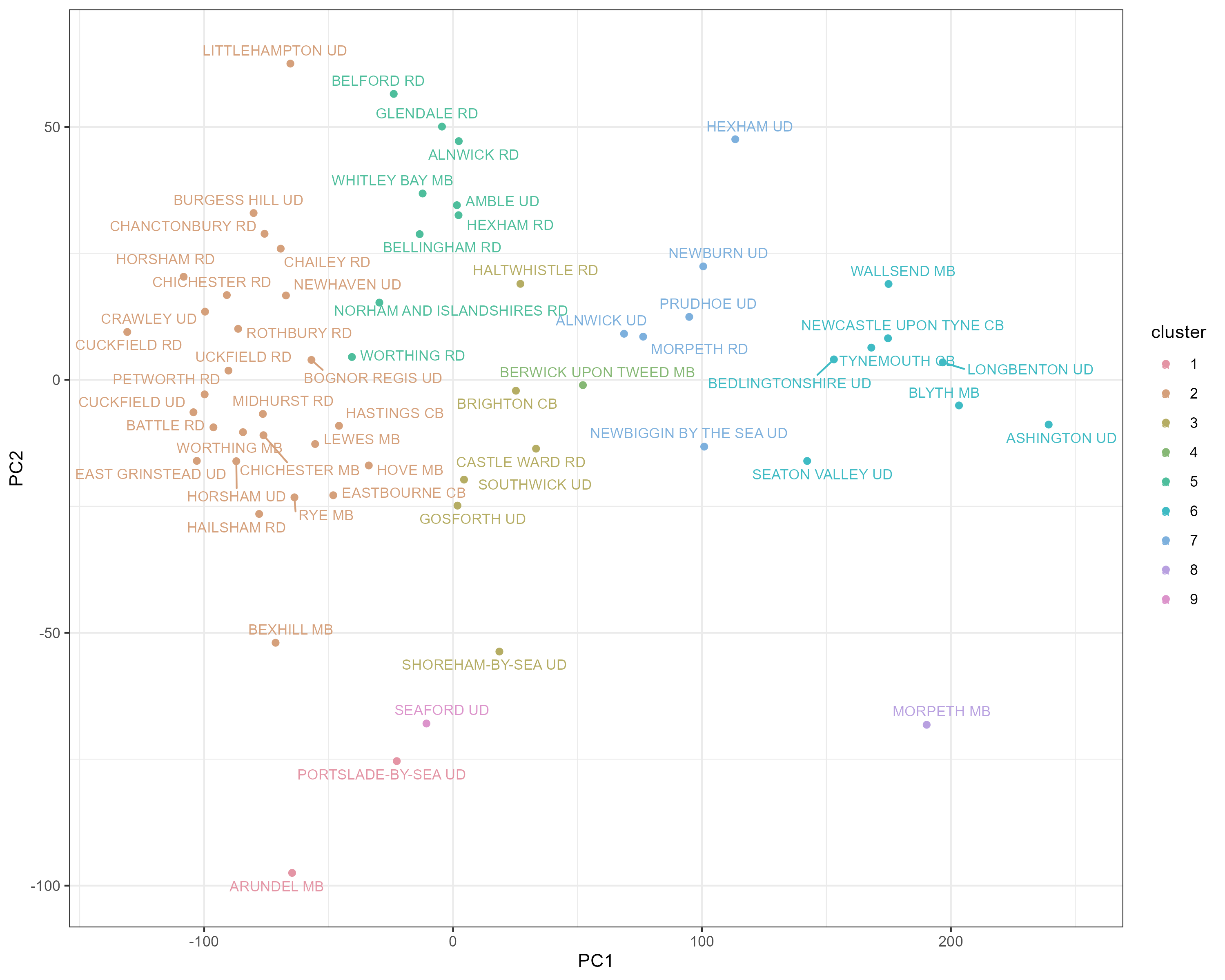}
\end{minipage} 
\hfill
\begin{minipage}
    {0.49\linewidth}
     \includegraphics[width=0.99\linewidth]{images/scatter_post_pc1_and_2_BRIGHTON.png}
\end{minipage} 
\caption{Scatter plot of functional principal component scores $1$ and $2$ --- local government districts plotted by cluster. Left: before adjusting for errors. Right: after adjusting for errors.}
\label{fig: hist cluster scatter plots pc1 and 2 both}
\end{figure} 

\begin{figure}[h!]
\centering
\begin{minipage}{0.49\linewidth}
     \includegraphics[width=0.99\linewidth]{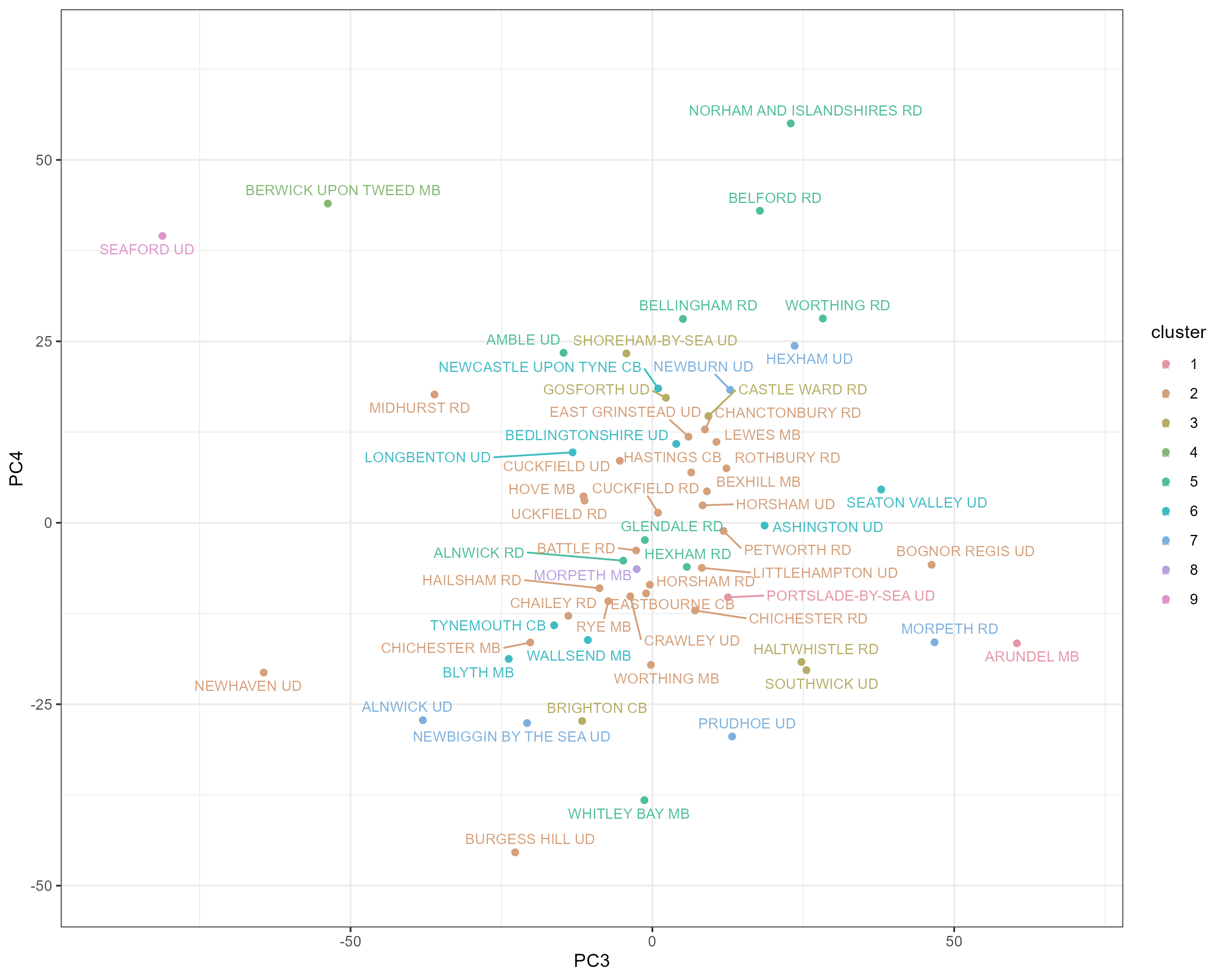}
\end{minipage} 
\hfill
\begin{minipage}
    {0.49\linewidth}
     \includegraphics[width=0.99\linewidth]{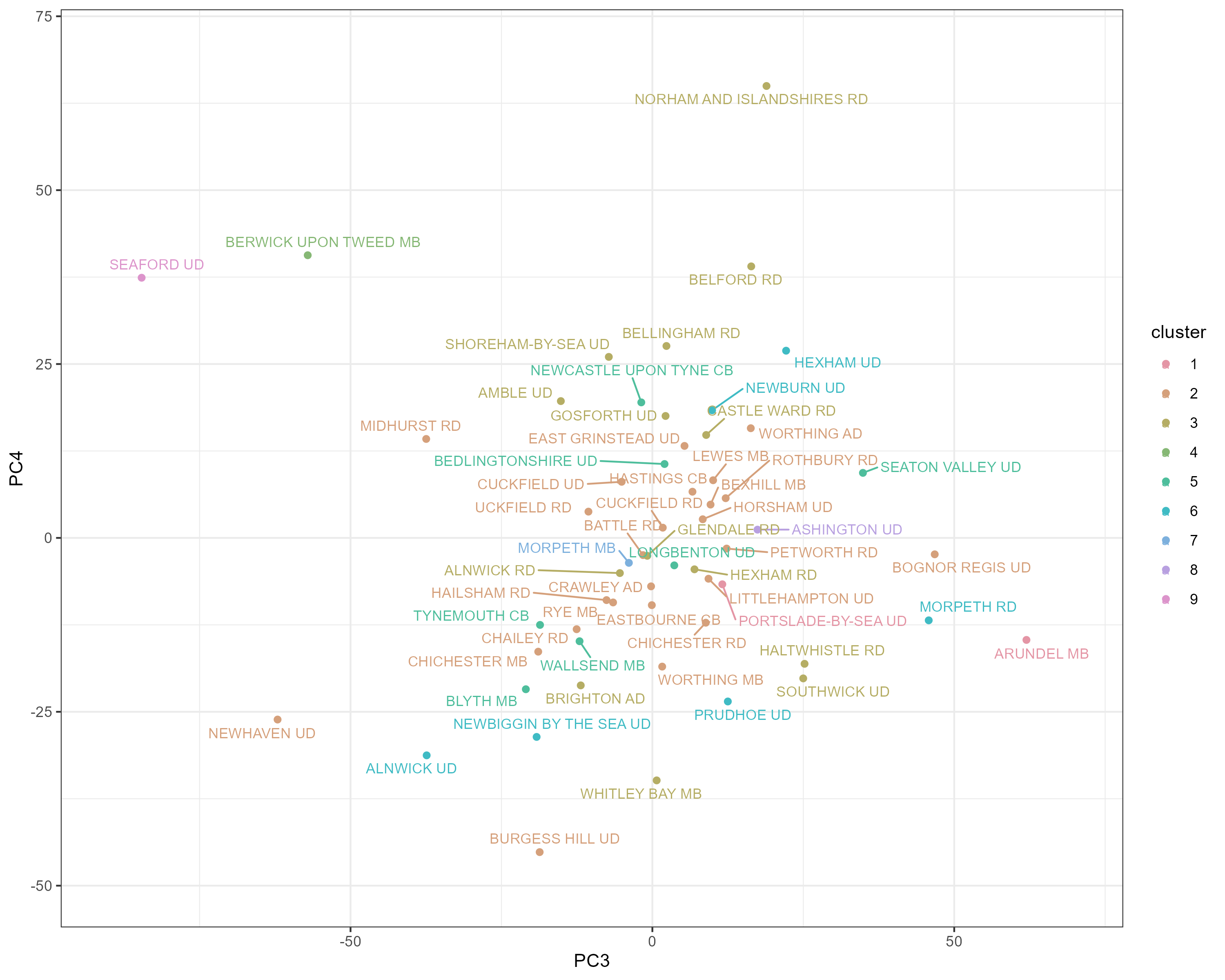}
\end{minipage} 
\caption{Scatter plot of functional principal component scores $3$ and $4$ --- local government districts plotted by cluster. Left: before adjusting for errors. Right: after adjusting for errors.}
\label{fig: hist cluster scatter plots pc3 and 4 both}
\end{figure} 

\end{appendices}

\ifOUP
  \section{Competing interests}
No competing interest is declared.



\section{Acknowledgments}

The infant mortality data analysed in this paper is from the Great Britain Historical Database (GBHDB), held by the UK Data Service,
\url{DOI: http://doi.org/10.5255/UKDA-SN-9035-1}, \copyright Southall, H.R., University of Portsmouth, Mooney, G., University of Portsmouth. 
The GIS files containing digital boundaries of local government districts $1911$-$1971$ are GBHDB
\url{ DOI: http://doi.org/10.5255/UKDA-SN-9321-1} \copyright Southall, H.R., University of Portsmouth, Gregory, I., Lancaster University, Aucott, P., University of Portsmouth, Burton, N., University of Portsmouth. 
These data sets are openly licensed via \url{https://creativecommons.org/licenses/by-sa/4.0/}. 
Computing in this paper is done in \texttt{R} \citep{Rproj}. Plots were produced using \texttt{ggplot2} \citep{wickham2011ggplot2}, except Figures \ref{fig: hist B-spline smoothed series} and \ref{fig: hist first four prin comp} which were produced using the \texttt{fda} package.

This paper is based on work completed while Tessa Wilkie was part of the EPSRC funded STOR-i Centre for Doctoral Training EP/S022252/1. She would also like to acknowledge the financial support of Shell Research Ltd during her PhD. She would like to thank Edward Austin for insightful discussions on the work in this paper. 
\fi

\bibliographystyle{abbrvnat}
\bibliography{reference}

\end{document}